\documentclass[prc,twocolumn,superscriptaddress,nofootinbib,
tightenlines,preprintnumbers,showkeys]{revtex4-1}

\usepackage[utf8]{inputenc}
\usepackage[english]{babel}
\usepackage{amssymb,amsthm,amsmath,amstext,amsbsy,amsopn}
\usepackage{bbm}
\usepackage{nicefrac}
\usepackage{slashed}
\usepackage{graphicx}
\usepackage{hyperref}
\usepackage{leftidx}
\usepackage{environ}
\usepackage{mathtools}
\usepackage{xspace}
\usepackage{array}
\usepackage{xcolor}
\usepackage{isotope}

\newcommand{\ie}{{i.e.}\xspace}
\newcommand{\eg}{{e.g.}\xspace}

\newcommand{\via}{{via}\xspace}

\newcommand{\mathspace}{\ \ }
\newcommand{\mathtext}[1]{\mathspace\text{#1}\mathspace}

\newcommand{\MeV}{\ensuremath{\mathrm{MeV}}}
\newcommand{\fm}{\ensuremath{\mathrm{fm}}}

\newcommand{\vecr}{\mathbf{r}}
\newcommand{\vecx}{\mathbf{x}}

\newcommand{\dd}{\mathrm{d}}

\newcommand{\del}[1]{\frac{\partial}{\partial #1}}
\newcommand{\deli}[2]{\frac{\partial^{#2}}{\partial #1^{#2}}}

\newcommand{\ii}{\mathrm{i}}

\newcommand{\OO}{\mathcal{O}}

\newcommand{\one}{\mathbbm{1}}

\newcommand{\sgn}{\mathrm{sgn}}

\newcommand{\ket}[1]{|#1\rangle}

\newcommand{\braket}[2]{\langle #1|#2\rangle}

\newcommand{\mbraket}[3]{\langle #1|#2|#3\rangle}

\newcommand{\abs}[1]{\left|#1\right|}

\newcommand*\rvec[1]%
{\ensuremath{\overset{\smash{\raisebox{-1.5pt}{\tiny$\rightarrow$}}}{#1}}}
\newcommand*\lvec[1]%
{\ensuremath{\overset{\smash{\raisebox{-1.5pt}{\tiny$\leftarrow$}}}{#1}}}

\NewEnviron{subalign}[1][]{%
\begin{subequations}\begin{align}
  \BODY
\end{align}\label{#1}\end{subequations}
}

\NewEnviron{spliteq}{%
\begin{equation}\begin{split}
  \BODY
\end{split}\end{equation}
}

\newcolumntype{K}[1]{>{\centering\arraybackslash}p{#1}}

\newcommand{\dvrsum}[2]{\sum\limits_{#1={-}#2/2}^{#2/2-1}}
\newcommand{\ophat}[1]{#1}

\begin{document}

\title{Signatures of few-body resonances in finite volume}

\author{P.~Klos}
\email{pklos@theorie.ikp.physik.tu-darmstadt.de}
\affiliation{Institut für Kernphysik, Technische Universität Darmstadt,
64289 Darmstadt, Germany}
\affiliation{ExtreMe Matter Institute EMMI,
GSI Helmholtzzentrum für Schwerionenforschung GmbH,
64291 Darmstadt, Germany}

\author{S.~König}
\email{sekoenig@theorie.ikp.physik.tu-darmstadt.de}
\affiliation{Institut für Kernphysik, Technische Universität Darmstadt,
64289 Darmstadt, Germany}
\affiliation{ExtreMe Matter Institute EMMI,
GSI Helmholtzzentrum für Schwerionenforschung GmbH,
64291 Darmstadt, Germany}

\author{H.-W.~Hammer}
\email{hans-werner.hammer@physik.tu-darmstadt.de}
\affiliation{Institut für Kernphysik, Technische Universität Darmstadt,
64289 Darmstadt, Germany}
\affiliation{ExtreMe Matter Institute EMMI,
GSI Helmholtzzentrum für Schwerionenforschung GmbH,
64291 Darmstadt, Germany}

\author{J.~E.~Lynn}
\email{joel.lynn@physik.tu-darmstadt.de}
\affiliation{Institut für Kernphysik, Technische Universität Darmstadt,
64289 Darmstadt, Germany}
\affiliation{ExtreMe Matter Institute EMMI,
GSI Helmholtzzentrum für Schwerionenforschung GmbH,
64291 Darmstadt, Germany}

\author{A.~Schwenk}
\email{schwenk@physik.tu-darmstadt.de}
\affiliation{Institut für Kernphysik, Technische Universität Darmstadt,
64289 Darmstadt, Germany}
\affiliation{ExtreMe Matter Institute EMMI,
GSI Helmholtzzentrum für Schwerionenforschung GmbH,
64291 Darmstadt, Germany}
\affiliation{Max-Planck-Institut f\"ur Kernphysik,
Saupfercheckweg 1,
69117 Heidelberg, Germany}

\begin{abstract}
We study systems of bosons and fermions in finite periodic boxes and show how
the existence and properties of few-body resonances can be extracted from
studying the volume dependence of the calculated energy spectra.  
We use and briefly review a plane-wave-based discrete variable
representation, which allows a convenient implementation of periodic boundary
conditions.  With these calculations we establish that avoided level crossings
occur in the spectra of up to four particles and can be linked to the existence 
of multibody resonances.  To benchmark our method we use two-body 
calculations, where resonance properties can be determined with other methods, 
as well as a three-boson model interaction known to generate a three-boson
resonance state.  Finding good agreement for these cases, we then predict
three-body and four-body resonances for models using a shifted Gaussian
potential.  Our results establish few-body finite-volume calculations as a new
tool to study few-body resonances.  In particular, the approach can be used to
study few-neutron systems, where such states have been conjectured to exist.
\end{abstract}

\maketitle

\section{Introduction}

The study of resonances, \ie, of short-lived, unstable states, constitutes
a very interesting and challenging aspect of few-body physics.  To
explore such systems theoretically, we discuss here the extraction of 
few-body-resonance properties from the volume dependence of energy levels in 
finite boxes with periodic boundary conditions.  Our study is motivated by 
recent efforts to observe~\cite{Kisa16tetraN,%
Pasc14rep,Kisa15rep,Shim15rep} and calculate~\cite{Wita993nRes,%
Laza053Nint,Laza05tetraNres,Hiya16tetraNres,Klos162n,Shir16tetraNres,%
Gand16EoTn,Foss17TetraN,Deltu183nRes} few-neutron resonances in
nuclear physics, but the scope is more general.

For two-body systems, it was shown by Lüscher~\cite{Lues86long,Lues91Torus}
that the infinite-volume properties of interacting particles are encoded in the
volume dependence of their (discrete) energy levels in the box.  These methods
are commonly used in the field of lattice
QCD~\cite{Bean11PPNP,Bric18lQCD}, but also in effective field theories (EFT)
with nucleon degrees of freedom~\cite{Bean04TwoNLat,Klos162n}.  The details of
extending the formalism from the two-body sector to few-body systems is a topic
of very active current research (see, \eg,
Refs.~\cite{Kreu11TriFV,Pole123pFV,Hans153pFV,Bric133pScat,%
Hamm173partFV,Koen17VDNB,Mai173pFV}).  In the two-particle sector, it was 
shown that a resonance leads to an avoided crossing of energy levels as the size 
$L$ of the box is varied~\cite{Wies89alc}.  This technique was used 
successfully to extract hadron resonances~(see Ref.~\cite{Bric18lQCD} for a 
recent review).  The same framework also applies to resonances in few-body 
systems which couple to an asymptotic two-body channel.

In the present work, we study the extension of this method to few-body
resonances.  In particular, we are interested in resonances that couple 
only to asymptotic three- or higher-body channels.  The properties of such 
systems, which one could refer to as ``genuine'' few-body resonances, cannot be 
obtained by calculating a standard two-body scattering phase shift.  Because to 
date there are no formal derivations for this case, we explore here whether such
states again show up as avoided crossings in the finite-volume few-body energy 
spectrum, and how the properties of the resonance state can be inferred from the 
position and shape of these avoided crossings.  Beyond establishing this method
as a tool for identifying resonance states, our results are relevant to test and 
help extend the ongoing formal work mentioned  above, in particular regarding 
the derivation of three-body finite-volume quantization
conditions~\cite{Pole123pFV,Hans153pFV,Bric133pScat,Hamm173partFV,%
Mai173pFV}.  We note that in a similar approach resonances can be studied in 
spherical boxes; see, for example, Refs.~\cite{Maie80spRes,Fedo09fbRes}.

Our studies require the calculation of several few-body energy levels in the
finite box.  An important consequence of the finite volume is that for any
given box size $L$ the spectrum is discrete, but it is still possible to
distinguish few-body bound states, which have an exponential volume
dependence~\cite{Koen17VDNB,Meis153bb}.  In contrast, continuum scattering 
states have a power-law volume dependence.  Resonances are
then identified as avoided crossings between these discrete ``scattering''
states as $L$ is varied (although we emphasize already here that in general
this signature is expected to be necessary, but not sufficient, for the
existence of resonance states).

Naturally, such calculations are numerically challenging, in particular
when the number of particles, the number of desired energy levels, or the
size of the volume increases.  As numerical method we use a discrete variable
representation~(DVR) based on an underlying basis of plane-wave eigenstates of
the box, which was previously applied to study few-nucleon systems 
in Ref.~\cite{Bulg13DVR}.  The latter allow one to conveniently 
implement periodic boundary conditions and naturally describe scattering states, 
and the use of the DVR promises significant advantages in computational 
efficiency over other methods~\cite{Groe01DVR,Bulg13DVR}.  We have developed a 
DVR framework that solves the finite-volume problem for both few-fermion and 
few-boson systems, supporting both small-scale (running on standard computers) 
as well as
efficient large-scale (running on high-performance computing clusters)
calculations.  An important challenge is to extend the reach of our method to
the very large box sizes that are required to unambiguously identify the
existence of proposed three- and four-neutron resonances at very low energies.
Postponing studies of few-neutron systems using EFT-based interactions to
future work, we investigate here systems of three and four bosons and fermions
using different model interactions.

This paper is organized as follows.  In Sec.~\ref{sec:num} we present the DVR
method applied to finite periodic boxes for both bosons and fermions,
discussing in some detail our numerical implementation.  This also addresses
the fact that in the periodic box one has to account for the breaking of
rotational symmetry to the cubic group, some details of which are given in the
appendix.  After discussing signatures of two-body resonances
in Sec.~\ref{sec:ResSig}, we proceed to the multibody case in
Sec.~\ref{sec:Res-34}, establishing first the validity of our approach using a
known three-body test case before we study bosonic and fermionic multibody
resonances using shifted Gaussian potentials.  We conclude in
Sec.~\ref{sec:Conclusion} with a brief summary and outlook.

\section{Numerical method}
\label{sec:num}

\subsection{Discrete variable representation}

To avoid contributions from the center-of-mass motion to the energy of
the system, we consider the $n$-body system in $n{-}1$ relative coordinates,
$\vecx_i = \vecr_n - \vecr_i$ for $i=1,\ldots,n{-}1$, where $\vecr_i$ denotes
the position of the $i$th particle.  These are not Jacobi coordinates, so
the kinetic energy operator $\ophat{T}_{\text{rel}}$ contains mixed derivatives 
in the position representation.  Because such terms are straightforward to deal 
with in the DVR representation, our choice of coordinates is convenient as 
it keeps the boundary conditions simple.  While the three-dimensional case is
physically the most relevant one, the construction here is completely general.
In $d$ spatial dimensions, the only difference is that all vectors have $d$
components.

\subsubsection{One-dimensional case}
\label{sec:DVR-1D}

The basic discussion of the DVR method given here follows that of
Ref.~\cite{Groe01DVR}, to which we also refer for more details.  To explain the
DVR method, we first consider two particles (with equal mass $m$ and reduced
mass $\mu = m/2$) in one spatial dimension, setting $x = \vecx_1$.  Confined to 
an interval of length $L$, periodic boundary conditions are imposed by choosing 
a basis of plane waves,
\begin{equation}
 \phi_j(x) = \braket{x}{\phi_j}
 = \frac{1}{\sqrt{L}}\exp(\ii p_j x) \,,
 \mathtext{with}
 p_j = \frac{2\pi j}{L} \,,
\label{eq:phi}
\end{equation}
and $i={-}N/2,\ldots,N/2-1$ with a truncation parameter $N$ (even) determining
the basis size.  It is clear that any periodic solution of the Schrödinger
equation,
\begin{equation}
 \left[\ophat{T}_{\text{rel}} + \ophat{V}\right] \ket{\psi} = E \ket{\psi}\,,
\label{eq:SG}
\end{equation}
can be expanded in the basis~\eqref{eq:phi}, and this representation becomes
exact for $N\to\infty$.

Following the DVR construction laid out in Ref.~\cite{Groe01DVR}, we consider
now pairs $(x_k, w_k)$ of grid points $x_k$ and associated weights $w_k$ such
that
\begin{equation}
 \dvrsum{k}{N} w_k\,\phi_i^*(x_k) \phi_j(x_k) = \delta_{ij} \,.
\label{eq:phi-orth}
\end{equation}
For the plane-wave basis~\eqref{eq:phi}, this is obviously satisfied by
\begin{equation}
 x_k = \frac{L}{N}k
 \mathtext{and}
 w_k = \frac{L}{N} \,.
\label{eq:xk-wk}
\end{equation}
If we now define matrices
\begin{equation}
 \mathcal{U}_{ki} = \sqrt{w_k} \phi_i(x_k) \,,
\label{eq:U-DVR}
\end{equation}
then these are unitary according to Eq.~\eqref{eq:phi-orth}, and we obtain the
DVR basis functions $\psi_k(x)$ by rotating the original plane-wave states:
\begin{equation}
 \psi_k(x) = \dvrsum{i}{N} \mathcal{U}^*_{ki} \phi_i(x)
\end{equation}
for $k = {-}N/2,\ldots,N/2-1$.  The range of indices is the same as
for the original plane-wave states, but whereas in Eq.~\eqref{eq:phi} they
specify a momentum mode, $\psi_k(x)$ is peaked at position $x_k \in
[{-}L/2,L/2)$.

It follows directly from Eqs.~\eqref{eq:xk-wk} and~\eqref{eq:U-DVR} as well as the
transpose $\mathcal{U}^T$ also being unitary that the DVR states have 
the property
\begin{equation}
 \psi_k(x_j) = \frac{1}{\sqrt{w_k}} \delta_{kj} \,.
 \label{eq:delta}
\end{equation}
This greatly simplifies the evaluation of the potential matrix elements:
\begin{spliteq}
 \mbraket{\psi_k}{\ophat{V}}{\psi_l}
 &= \int \dd x\,\psi_k^*(x) V(x) \psi_l(x)\,, \\
 &\approx \dvrsum{m}{N} w_m\,\psi_k^*(x_m) V(x_m) \psi_l(x_m)\,, \\
 &= V(x_k) \delta_{kl} \,,
\label{eq:V-DVR}
\end{spliteq}
so that the potential operator is (approximately) diagonal in the DVR
representation.  The approximation here lies in the second step in
Eq.~\eqref{eq:V-DVR}, replacing the integral by a sum, which is possible because
the $(x_k,w_k)$ defined in Eq.~\eqref{eq:xk-wk} constitute the mesh points and
weights of a trapezoidal quadrature rule.  Note that for this
identification it is important that the points ${-}L/2$ and $L/2$ are identified
through the periodic boundary condition because otherwise the weight
$w_{{-}N/2}$ would be incorrect.

The kinetic energy, given in configuration space by the differential operator
\begin{equation}
 T_{\text{rel}} = {-}\frac{1}{2\mu}\frac{\dd^2}{\dd x^2} \,,
\label{eq:T-rel-1D}
\end{equation}
is not diagonal in the DVR representation (note that we set
here $\hbar = 1$).  However, its matrix elements can be written in closed
form~\cite{Bulg13DVR}:
\begin{equation}
 \mbraket{\psi_k}{\ophat{T}_{\text{rel}}}{\psi_l} = \begin{cases}
  \dfrac{\pi^2 N^2}{6\mu L^2} \left(1+\dfrac{2}{N^2}\right)\,,
  &\text{for}\quad k = l\,, \\[10pt]
  \dfrac{({-}1)^{k-l} \pi^2}{\mu L^2 \sin^2\big(\pi(k-l)/N\big)}\,,
  &\text{otherwise}\,.
 \end{cases}
\label{eq:T-DVR-1D}
\end{equation}
While this matrix is dense here, we will see below that it becomes sparse for
$d > 1$.  Alternatively, as pointed out in Ref.~\cite{Bulg13DVR}, one can
use a discrete fast Fourier transform to evaluate the kinetic energy in
momentum space.  This operation switches from the DVR to the original
plane-wave basis~\eqref{eq:phi}, where we have
\begin{equation}
 \ophat{T}_{\text{rel}}\ket{\phi_i} = \frac{p_i^2}{2\mu}\ket{\phi_i} \,,
\end{equation}
and back.

\subsubsection{General construction}
\label{sec:DVR-General}

The construction is straightforward to generalize to the case of an arbitrary
number of particles $n$ and spatial dimensions $d$: The starting point simply
becomes a product of $(n-1)\times d$ plane waves, one for each
relative-coordinate component. The transformation matrices and DVR basis
functions are defined \via tensor products.  Eventually, while a single index
suffices to label the one-dimensional DVR states, a collection of
$(n-1)\times d$ indices defines the general case.  For these states we
introduce the notation (generalizing the 1D short-hand form
$\ket{\psi_k} = \ket{k}$)
\begin{equation}
 \ket{s}
 = \ket{
  (k_{{1,1}},\cdots,k_{{1,d}}),\cdots,(k_{{n-1,1}},\cdots);
  (\sigma_1,\cdots,\sigma_n)
 } \,.
\label{eq:s}
\end{equation}
Here we have also included additional indices to account for spin degrees of
freedom.  If the particles have spin $S$, then each $\sigma_i$, labeling the
projections, takes values from ${-}S$ to $S$.  Additional internal
degrees of freedom, such as isospin, can be included in the same way.  The
collection of all these states $\ket{s}$ is denoted by $B$, which
is our DVR basis with dimension $\dim B = (2S+1)^n \times N^{(n-1)d}$.

We take the interaction $\ophat{V}$ in Eq.~\eqref{eq:SG} to be a sum of
central, local $A$-body potentials (with $A=2,\ldots,n$ for an $n$-body system).
Each contribution to this sum depends only on the relative distances 
between pairs of particles.  This means that matrix elements of $\ophat{V}$ 
between $n$-particle states depend on $n{-}1$ relative coordinates, and for each 
of these there is a delta function in the matrix element,
\begin{multline}
 \mbraket{\vecx_1,\cdots,\vecx_{n-1}}{\ophat V}
 {\vecx_1',\cdots,\vecx_{n-1}'} \\
 = V(\{\abs{\vecx_i}\},\{\abs{\vecx_{i}-\vecx_{j}}\}_{i< j})
 \prod_i \delta^{(d)}(\vecx_i'-\vecx_i) \,,
\label{eq:V-gen}
\end{multline}
so that the interaction remains diagonal in the general DVR basis.  For the
evaluation between DVR states $\ket{s}$, each modulus $\abs{\vecx_i}$ in
Eq.~\eqref{eq:V-gen} gets replaced with
\begin{equation}
 \abs{s_i}
 \equiv \frac{L}{N}\left(\sum\limits_{c=1}^d k_{i,c}^2\right)^{1/2} \,.
\end{equation}
If the potential depends on the spin degrees of freedom, the potential matrix
in our DVR representation acquires nondiagonal terms, but these are determined
solely by overlaps in the spin sector, and overall this matrix remains very
sparse.

As already pointed out, the kinetic energy matrix is also sparse in $d>1$.  To
see this, first note that the 1D matrix elements~\eqref{eq:T-DVR-1D} enter
for each component $k_{i,c}$, multiplied by Kronecker deltas for each
$c' \neq c$ and summed for all relative coordinates $i=1,\ldots,n-1$.
The only additional complication, stemming from our choice of simple relative
coordinates, is that the general kinetic energy operator,
\begin{equation}
 T_{\text{rel}}^{\text{$n$-body}}
 = {-}\frac{1}{2\mu} \sum\limits_{i=1}^{n-1}\sum\limits_{j=1}^i
 \del{x_i}\del{x_j} \,,
\label{eq:T-rel}
\end{equation}
contains mixed (non-diagonal) terms.  As an example to illustrate this, consider
the kinetic-energy operator for three particles in one dimension,
\begin{equation}
 T_{\text{rel}}^{\text{3-body}}
 =-\frac{1}{2\mu}\biggl(\deli{x_1}{2}+\deli{x_2}{2}+\del{x_1}\del{x_2}\biggr)
 \,.
\end{equation}
For this the kinetic-energy matrix elements are given by
\begin{multline}
 \mbraket{k_{1}k_{2}}{\ophat{T}_{\text{rel}}^{\text{3-body}}}{l_{1}l_{2}}
 = \mbraket{k_1}{\ophat{T}_{\text{rel};1}}{l_1}\delta_{k_2 l_2} \\
 + \mbraket{k_2}{\ophat{T}_{\text{rel};2}}{l_2}\delta_{k_1 l_1}
 + \mbraket{k_{1}k_{2}}{\ophat{T}_{\text{rel};12}}{l_{1}l_{2}} \,,
\end{multline}
where the first two matrix elements on the right-hand side are given in 
Eq.~\eqref{eq:T-DVR-1D} and the last term is a special case of the general 
mixed-derivative operator
\begin{equation}
 T_{\text{rel};ij}
 = {-}\frac{1}{2\mu} \del{x_{i}}\del{x_{j}} \,.
\end{equation}
The DVR matrix elements for this are given by
\begin{equation}
 \mbraket{k_{i}k_{j}}{\ophat{T}_{\text{rel};ij}}{l_{i}l_{j}}
 = {-}\frac{1}{2\mu}
 \big[\mbraket{k_i}{\partial_i}{l_i} \mbraket{k_j}{\partial_j}{l_j}\big]
\end{equation}
with~\cite{Bila17bsc}
\begin{equation}
 \mbraket{k}{\partial}{l} = \begin{cases}
  {-}\ii\dfrac{\pi}{L}\,,
  &\text{for}\quad k = l\,, \\
  \dfrac{\pi}{L}
  \dfrac{({-}1)^{k-l}\exp\!\left({-}\ii\dfrac{\pi(k-l)}{N}\right)}
  {\sin\!\left(\dfrac{\pi(k-l)}{N}\right)}\,,
  &\text{otherwise}\,.
 \end{cases}
\end{equation}
As for the diagonal terms, for a general state $\ket{s}$ these terms are
summed over for all pairs of relative coordinates and spatial components $c$,
including Kronecker deltas for $c' \neq c$.

Analogous to the one-dimensional case the kinetic energy can alternatively be
implemented by switching to momentum space with a fast Fourier transform,
applying a diagonal matrix with entries
\begin{equation}
 T_{\text{rel}}^{\text{$n$-body}}\ket{s} = \frac{1}{2\mu L^2}
 \sum\limits_{i=1}^{n-1}\sum\limits_{j=1}^i\sum_{c=1}^d
 k_{i,c} k_{j,c}\ket{s} \,,
\label{eq:T-rel-s}
\end{equation}
and then transforming back with the inverse transform.

\subsubsection{(Anti-)symmetrization and parity}
\label{sec:AntiSymm-Parity}

To study systems of identical bosons (fermions), we want to consider
(anti-)symmetrized DVR states.  The construction of these can be achieved with
the method described, \eg, in Ref.~\cite{Varg97fb} (for the stochastic
variational model in Jacobi coordinates):
\begin{enumerate}
\item The transformation from single-particle to relative coordinates is
written in matrix form as
\begin{equation}
 \vecx_i = \sum\limits_{j=1}^n U_{ij} \vecr_j \,,
\end{equation}
where
\begin{equation}
 U_{ij} = \begin{cases}
  \delta_{ij}\,, &\text{for}\quad i,j < n\,, \\
  {-1}\,,        &\text{for}\quad i < n,\, j = n\,, \\
  1/n\,,         &\text{for}\quad i = n\,.
 \end{cases}
\end{equation}
Note that for $i=n$ this definition includes the center-of-mass coordinate.
\item For the $n$-particle system there are $n!$ permutations, constituting the
symmetric group $S_n$.  A permutation $p \in S_n$ can be represented as a
matrix $C(p)$ with
\begin{equation}
 C(p)_{ij} = \begin{cases}
  1\,, &\text{for}\quad j = p(i)\,, \\
  0\,, &\text{otherwise}\,,
\end{cases}
\end{equation}
acting on the single-particle coordinates $\vecr_i$.
\item The operation of $p \in S_n$ on the relative coordinates is then given by
the matrix
\begin{equation}
 C_{\text{rel}}(p) = U\,C(p)\,U^{{-1}} \,,
\end{equation}
with the row and column of the left-hand side discarded, so that
$C_{\text{rel}}(p)$ is an $(n{-}1)\times(n{-}1)$ matrix.
\end{enumerate}

Because the indices $k_{i,c}$ correspond directly to positions on the spatial grid
via Eq.~\eqref{eq:delta}, acting with $C_{\text{rel}}(p)$ on a state $\ket{s}$
is now straightforward: The $k_{i,c}$ are transformed according to the entries
$C_{\text{rel}}(p)_{ij}$, where for each $i$ one considers all $c=1,\ldots,d$ at
once. In other words, $C_{\text{rel}}(p)$ is expanded (by replication for each
$c$) to a matrix acting in the space of individual coordinate components.
As a final step, to maintain periodic boundary conditions, any transformed
indices that may fall outside the original range ${-}N/2,\ldots,N/2-1$ are
wrapped back into this interval by adding appropriate multiples of $N$.
Applying the permutation to the spin indices $(\sigma_1,\ldots,\sigma_n)$ is
trivial because they are given directly as an $n$-tuple.  The final result of
this process for a given state $\ket{s} \in B$ and permutation $p$ is a
transformed state,
\begin{equation}
 \ket{s'} = \mathcal{C}(p) \ket{s} \in B \,,
\label{eq:C-p}
\end{equation}
where
\begin{equation}
 \mathcal{C}(p) = C_{\text{rel}}(p) \, C_{\text{spin}}(p)
\end{equation}
denotes the total permutation operator in the space of DVR states.  The
statement of Eq.~\eqref{eq:C-p} is that each $p\in S_n$ acts on $B$ as a whole
by permuting the order of elements.

With this, we can now define the symmetrization and antisymmetrization
operators as
\begin{equation}
 \mathcal{S} = \frac{1}{n!}\sum_{p\in S_n} \mathcal{C}(p)
 \mathtext{and}
 \mathcal{A} = \frac{1}{n!}\sum_{p\in S_n} \sgn(p)\,\mathcal{C}(p) \,,
\end{equation}
where $\sgn(p) = \pm1$ denotes the parity of the permutation $p$.  Because both of
these operators are projections ($\mathcal{S}^2 = \mathcal{S}$, $\mathcal{A}^2 =
\mathcal{A}$), they map our original basis $B$ onto bases $B_\mathcal{S/A}$
of, respectively, symmetrized or antisymmetrized states, each consisting of 
linear combinations of states in $B$.  An important feature of these mappings is
that each $\ket{s}\in B$ appears in at most one state in
$B_\mathcal{S}$ (for symmetrization) or $B_\mathcal{A}$ (for 
antisymmetrization).  Thus, to determine $B_\mathcal{S}$ we
can simply apply $\mathcal{S}$ to all $\ket{s}\in B$, dropping duplicates, and 
analogously for the construction of $B_\mathcal{A}$.  Moreover, for the
practical numerical implementation of this procedure (discussed in more detail 
in Sec.~\ref{sec:Numerics-Details}) it suffices to store a single term for each 
linear combination because the full state can be reconstructed from that through 
an application of the (anti-)symmetrization operator.

Parity can be dealt with in much the same way: The parity operator $\mathcal{P}$
merely changes the sign of each relative coordinate, so it can be applied to the
DVR states defined in Eq.~\eqref{eq:s} by mapping $k_{i,c} \to {-}k_{i,c}$ for
all $i,c$, and, if necessary, wrapping the result back into the range
${-}N/2,\ldots,N/2-1$.  The spin part remains unaffected by this operation.
Projectors onto positive and negative parity states are given as
\begin{equation}
 \mathcal{P}_\pm = \one \pm \mathcal{P} \,.
\label{eq:P-pm}
\end{equation}
They have the same properties as $\mathcal{S}$ and $\mathcal{A}$ (each
$\ket{s}\in B$ appears in at most one linear combination forming a state with
definite parity), and, importantly, the same is true for the combined operations
$\mathcal{P}_\pm\mathcal{S}$ and $\mathcal{P}_\pm\mathcal{A}$.  In practice this
means that it is possible to efficiently construct bases of (anti-)symmetrized
states with definite parity, where for each element it suffices to know a single
generating element $\ket{s}\in B$.

\subsubsection{Cubic symmetry projection}
\label{sec:Cubic}

While permutation symmetry and parity remain unaffected by the finite periodic
geometry, rotational symmetry is lost.  In particular, in $d = 3$ dimensions (to
which the remaining discussion in this subsection will be limited), angular
momentum $l$ is no longer a good quantum number for the $n$-body system in the
periodic cubic box.  Specifically, the spherical $SO(3)$ symmetry of the 
infinite-volume system is broken down to a cubic subgroup $\OO \subset SO(3)$.

This group has 24 elements and five irreducible representations $\Gamma$,
conventionally labeled $A_1$, $A_2$, $E$, $T_1$, and $T_2$.  Their
dimensionalities are $1$, $1$, $2$, $3$, and $3$, respectively, and irreducible
representations $D^l$ of $SO(3)$, determining angular-momentum multiplets in
the infinite volume, are reducible with respect to $\OO$.  As a result, a given 
(infinite-volume) angular momentum state can contribute to several $\Gamma$.
In the cubic finite volume, one finds the spectrum decomposed into multiplets 
with definite $\Gamma$, where an index $\alpha=1,\ldots,\dim\Gamma$ further 
labels the states within a given multiplet.

For our calculations, it is desirable to select spectra by their cubic
transformation properties.  To that end, we construct projection
operators~\cite{John82AML},
\begin{equation}
 \mathcal{P}_\Gamma
 = \frac{\dim\Gamma}{24} \sum_{R\in\OO}\,\chi_\Gamma(R) D_{n}(R) \,,
\label{eq:P_Gamma}
\end{equation}
where $\chi_\Gamma(R)$ denotes the character (tabulated in
Ref.~\cite{John82AML}) of the cubic rotation $R$ for the irreducible
representation $\Gamma$ and $D_{n}(R)$ is the realization of the cubic rotation
in our DVR space of periodic $n$-body states.  For example, for the 
one-dimensional representation $\Gamma=A_1$, $\chi_{A_1}(R)=1$ for all cubic 
rotations $R$, so in this case Eq.~\eqref{eq:P_Gamma} reduces to an average 
over all rotated states.  In Appendix~\ref{app:cubic} we provide some further 
discussion of the cubic group and the construction of the $D_{n}(R)$.

\subsection{Implementation details}
\label{sec:Numerics-Details}

We use a numerical implementation of the method described above written
predominantly in C++, with some smaller parts (dealing with permutations)
conveniently implemented in Haskell.  For optimal performance, parallelism \via
threading is used wherever possible.  Our design choice to use modern C++11
allows us to achieve this by means of the TBB library~\cite{TBB}, which provides
high-level constructs for nested parallelism as well as convenient concurrent
data structures.  To support large-scale applications, we also split
calculations across multiple nodes using MPI, so that overall we have a hybrid
parallel framework.

For a fixed setup (given physical system, box size $L$, DVR truncation
parameter $N$), the calculation is divided into three phases:
\begin{enumerate}
\item Basis setup
\item Hamiltonian setup
\item Diagonalization
\end{enumerate}
The last step is the simplest one conceptually, so we start the discussion from
that end.  To calculate a given number of lowest energy eigenvalues 
we use the parallel ARPACK package~\cite{PARPACK-ng}, implementing 
Arnoldi/Lanczos iterations distributed via MPI.  This method requires the
calculation of a number of matrix-vector products,
\begin{equation}
 \psi_{\text{out}} = \ophat{H} \psi_{\text{in}} \,,
\label{eq:psi-in-out}
\end{equation}
applying the DVR Hamiltonian $\ophat{H}$ to state vectors
$\psi_{\text{in}}$ (provided by the algorithm) until convergence is
reached.  These are potentially very large (see Sec.~\ref{sec:DVR-General})
and thus are distributed across multiple nodes.  Explicit synchronization is
only required for $\psi_{\text{in}}$ to evaluate the right-hand
side of Eq.~\eqref{eq:psi-in-out}. Each node only calculates its local
contribution to $\psi_{\text{out}}$.

We note here that while (anti-)symmetrization and parity are directly
realized by considering appropriate basis states, the simplifications discussed
in Sec.~\ref{sec:AntiSymm-Parity} are not possible for the cubic-symmetry
projectors $\mathcal{P}_\Gamma$ introduced in Sec.~\ref{sec:Cubic}.  Instead,
the latter are accounted for \via the substitution,
\begin{equation}
 \ophat{H} \to \ophat{H} + \lambda(\one - \mathcal{P}_\Gamma) \,,
\end{equation}
where $\lambda$ is an energy scale chosen much larger than the energy of the
states of interest.  This construction applies a shift to all states which do
not possess the desired symmetry, leaving only those of interest in the
low-energy spectrum obtained with the Lanczos algorithm.

The operator $\mathcal{P}_\Gamma$ is constructed as a large sparse matrix,
which we implement using Intel MKL~\cite{MKL}, if available, and via
\texttt{librsb}~\cite{librsb} otherwise.  The same holds for the kinetic-energy
matrix when operating in a mode where this matrix is constructed explicitly (as
described in Sec.~\ref{sec:DVR-General}) in step 2.

While this mode of operation has good scaling properties with increasing number
of compute nodes, we find it to be overall more efficient (in particular with
respect to the amount of required memory) to use the Fourier-transform-based
kinetic-energy application, which we implement using FFTW~\cite{FFTW}.  Because
the transform is defined for the full (not symmetry-reduced)
basis, this method involves transforming the vectors $\psi_{\text{in}}$ to the
large space, and transforming back after applying the kinetic-energy operator.
These transformations are again implemented \via sparse-matrix multiplications,
where the matrix $X$ that expands from the reduced space to the full space has
entries given by eigenvectors of (appropriate combinations of) the operators
$\mathcal{S}$, $\mathcal{A}$, and $\mathcal{P}_\pm$ described in
Sec.~\ref{sec:AntiSymm-Parity}.  Reducing back at the end is performed with the
transpose matrix $X^T$.  For calculations on multiple nodes using MPI,
individual ranks need only calculate local slices of these matrices.

In Fourier-transform mode, step 2 consists only of calculating diagonal
matrices for the kinetic energy and the potential parts of the Hamiltonian, and
possibly of setting up the sparse cubic projection matrix $\mathcal{P}_\Gamma$.
These calculations are based on determining the symmetry-reduced basis states
in step $1$, which can be efficiently parallelized across multiple nodes.  In
addition, this requires calculating $X$ and $X^T$.\footnote{On a single node, 
it is sufficient to calculate just one of these matrices.  For distributed
calculations, however, different nodes need different slices of these matrices
so that in order to reduce communication overhead it is most efficient to store
both $X$ and $X^T$.}

\section{Resonance signatures}
\label{sec:ResSig}

\begin{figure}[btp]
\centering
\includegraphics[width=0.99\columnwidth]{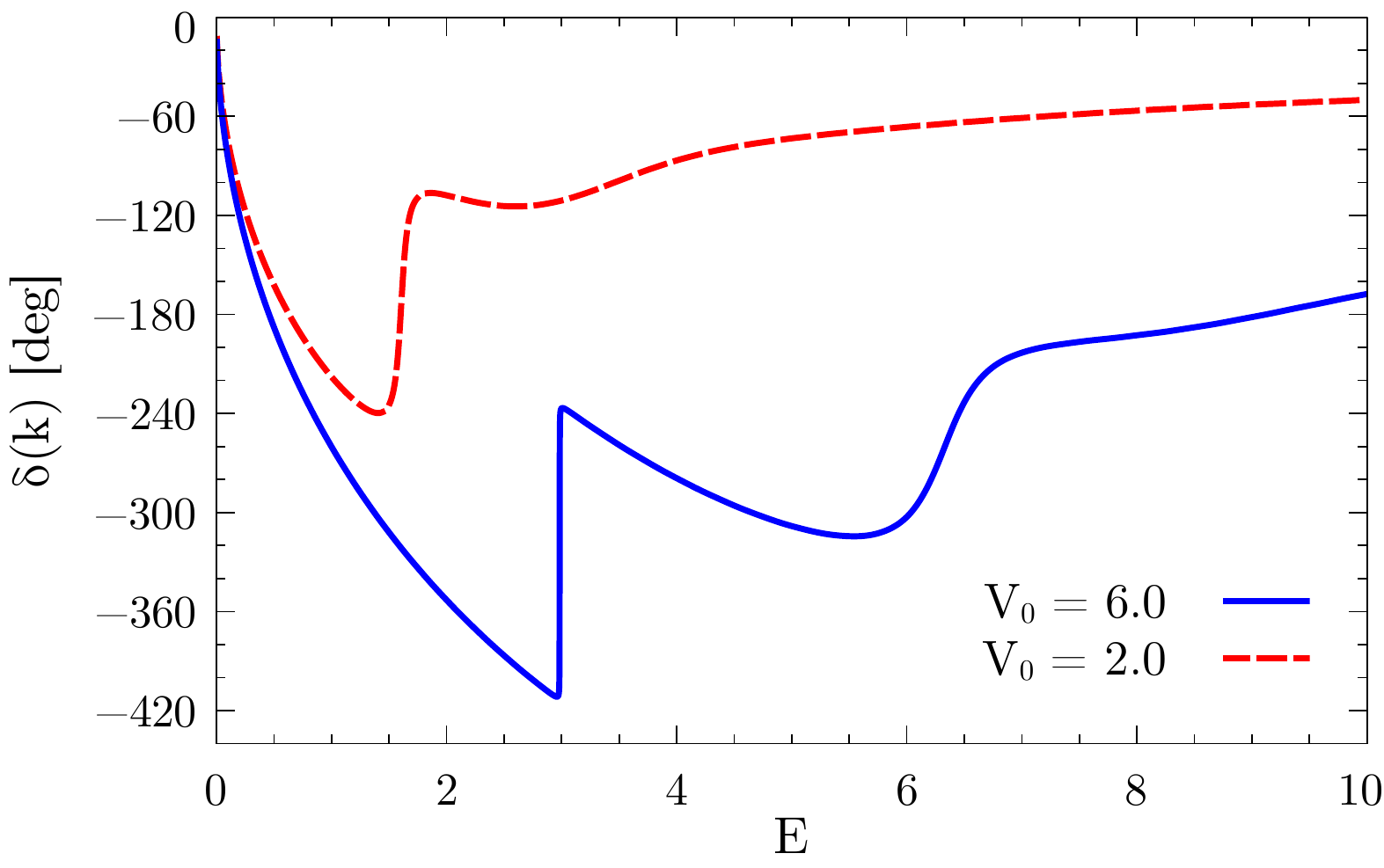}
\caption{$S$-wave phase shift of two particles interacting via the
potential given in Eq.~\eqref{eq:shiftedGauss} as a function of the
(dimensionless) relative kinetic energy $E$ for $V_0=6.0$ (blue solid curve)
and $V_0=2.0$ (red dashed curve).}
\label{fig:Phase-d3n2b}
\end{figure}

In the two-particle sector it has been shown that a resonance state manifests
itself as avoided level crossings when studying the volume dependence of the
discrete energy levels in a periodic box~\cite{Wies89alc}.  Before we move on to
establish the same kind of signature for more than two particles in the
following section, we compare here the finite-volume resonance determination to
other methods.  As a test case, we consider two particles interacting \via a
shifted Gaussian potential,
\begin{align}
 V(r)=V_0 \exp \biggl(-\Bigl(\frac{r-a}{R_0}\Bigr)^2\biggr) \,.
\label{eq:shiftedGauss}
\end{align}
This kind of repulsive barrier is very well suited to produce narrow resonance 
features without much need for fine tuning.  To illustrate this we show in
Fig.~\ref{fig:Phase-d3n2b} $S$-wave scattering phase shifts for $a=3$, $R_0=1.5$
and two different values of  $V_0$ (all in natural units, which besides using 
$\hbar=c=1$ also set $m=1$).  For $V_0=6.0$ the phase shift exhibits a very 
sharp jump of approximately $180^{\circ}$.  From the location of the inflection
point of the phase shift we extract the resonance energy $E_R$, while the width 
$\Gamma$ is given by the value of the derivative at the resonance energy,
\begin{equation}
 \biggl[\frac{\dd\delta(E)}{\dd E} \biggr]_{E=E_R}=\frac{2}{\Gamma} \,.
\end{equation}
We find a very narrow two-body resonance with energy $E_R=2.983$ and width
$\Gamma=0.001$.  When the height of the barrier is lowered to $V_0=2.0$, the 
jump is much less pronounced, implying that the width of this resonance is 
broadened.  Indeed, we find resonance parameters of $E_R=1.606$, $\Gamma=0.097$ 
for this case.

To further check these parameters, we consider Eq.~\eqref{eq:shiftedGauss}
Fourier transformed to momentum space and look for poles in the $S$-wave
projected $S$-matrix on the second energy sheet, using the technique described 
in Ref.~\cite{Gloeckle83FB}.  For $V_0=6.0$ we find a resonance pole at
$E_R-\ii\Gamma/2 = 2.9821(3) - \ii 0.00035(5)$, where the uncertainty is
estimated by comparing calculations with $300$ and $256$ points for a
discretized momentum grid with cutoff $8$ (in natural inverse length units).
In the same way, we extract $E_R-\ii\Gamma/2 = 1.606(1) - \ii
0.047(2)$ for $V_0=2.0$.  Noting that there is no completely unambiguous way
to relate the parameters extracted from the phase shifts (except in the limit
of vanishing background and poles infinitesimally close to the real axis), we
conclude that these pole positions are in very good agreement with the behavior
seen in the phase shifts.

%%%%%%%%%%%%%%%%%%%%%%%%%%%%%%%%%%%%%%%%%%%%%%%%%%%%%%%%%%%%%%%%%%%%%%%%%%%%%%
%%%%%%%%%%%%%%%%%%%%%%%%%%%%%%%%%%%%%%%%%%%%%%%%%%%%%%%%%%%%%%%%%%%%%%%%%%%%%%

%%%%%%%%%%%%%%%%%%%%%%%%%%%%%%%%%%%%%%%%%%%%%%%%%%%%%%%%%%%%%%%%%%%%%%%%%%%%%%
\begin{figure*}[htbp]
\centering
\begin{minipage}{0.5\textwidth}
\centering
\includegraphics[width=0.99\columnwidth]{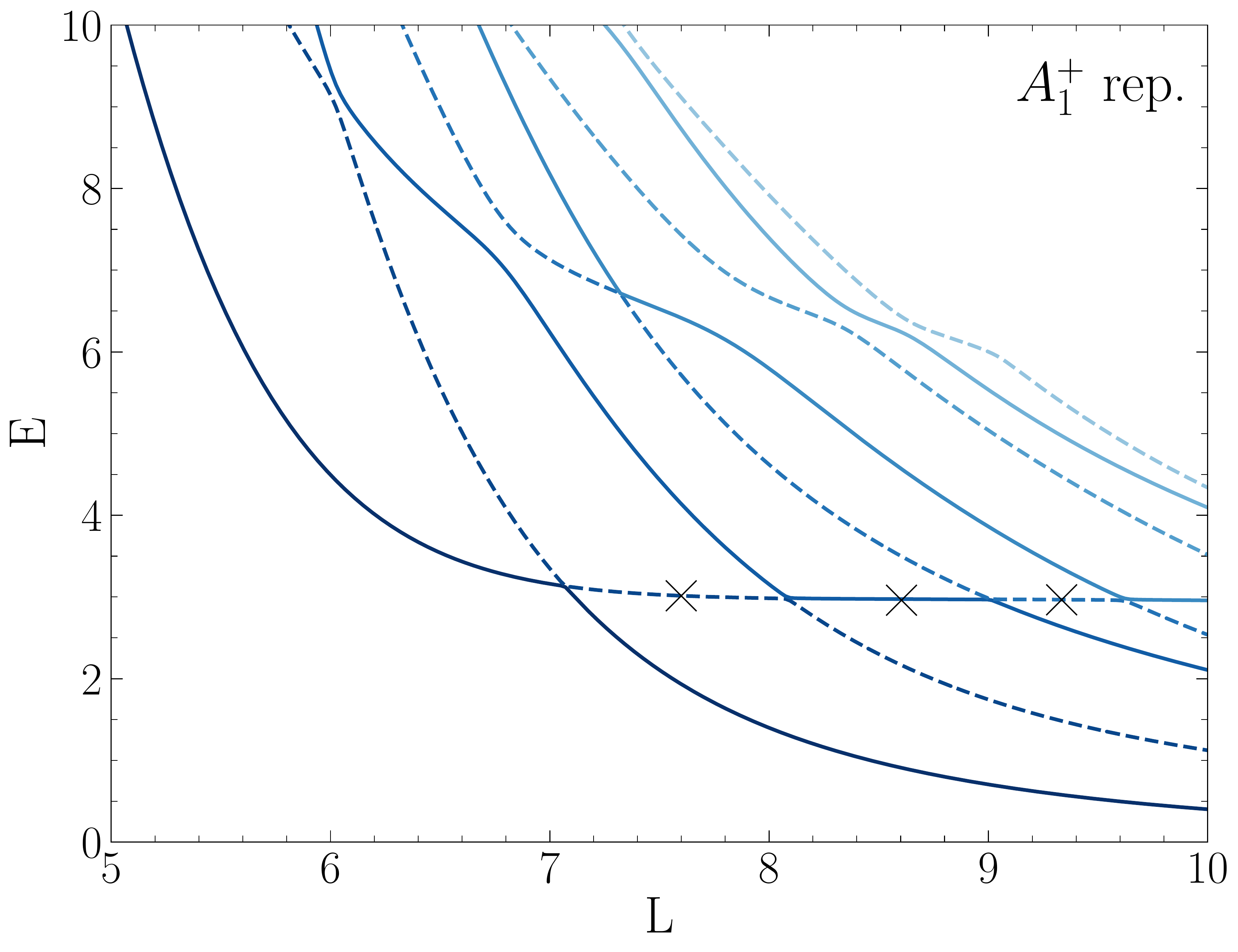}
\end{minipage}\begin{minipage}{0.5\textwidth}
\centering
\includegraphics[width=0.99\columnwidth]{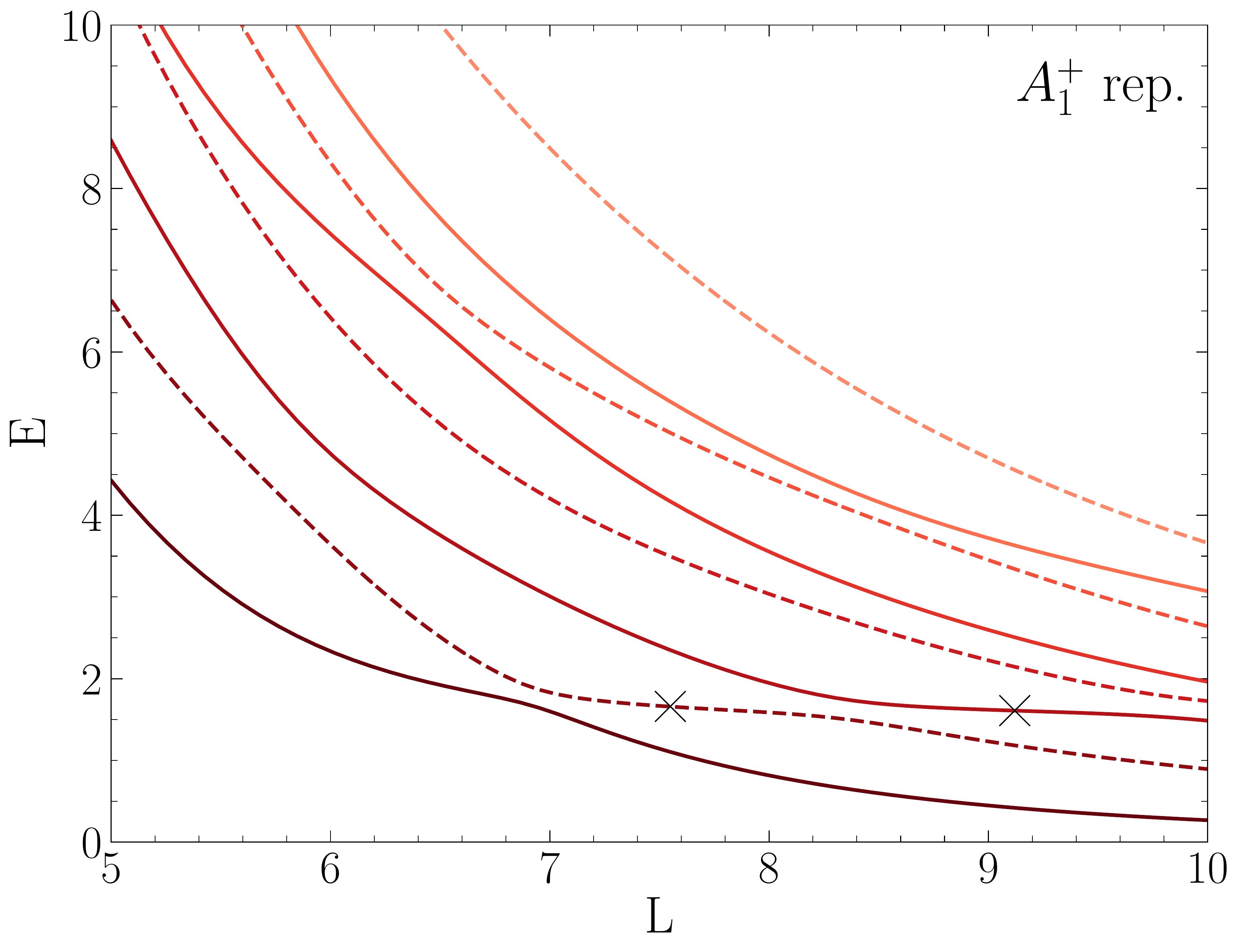}
\end{minipage}
\caption{Energy spectrum of two particles interacting \via the
potential given in Eq.~\eqref{eq:shiftedGauss} in finite volume for different 
box sizes $L$.  The left panel shows results for $V_0=6.0$ in the $A_1^+$ 
representation, whereas for the right panel a weaker barrier $V_0=2.0$ was 
used.  All crossings are avoided because the spectrum is fully 
projected on states with the same quantum numbers.  The crosses mark 
the inflection points used to extract the resonance energy (see text).}
\label{fig:En-2b-Dietz-A1p}
\end{figure*}
%%%%%%%%%%%%%%%%%%%%%%%%%%%%%%%%%%%%%%%%%%%%%%%%%%%%%%%%%%%%%%%%%%%%%%%%%%%%%%

We now perform finite-volume calculations of two particles in a
three-dimensional box with periodic boundary conditions using the DVR method
discussed in Sec.~\ref{sec:num}.  As avoided level crossings corresponding to a
resonance are only expected for states with the same quantum numbers, we
project onto states that belong to a single irreducible representation $\Gamma$
of the cubic group (see Sec.~\ref{sec:Cubic}) and definite parity. Specifically,
we consider here only $A_1^+$ states, which to a good approximation correspond
to $S$-wave states in the infinite volume.  As shown in
Table~\ref{tab:cubicL}, the next higher angular momentum contributing to $A_1^+$
is $l=4$, which can be safely neglected for low-energy states.

Our results are shown in Fig.~\ref{fig:En-2b-Dietz-A1p}.  In the spectrum for
$V_0=6.0$ (left panel of Fig.~\ref{fig:En-2b-Dietz-A1p}), a series of extremely
sharp avoided level crossings, forming an essentially horizontal plateau, is
observed at approximately $E \approx 3.0$.  According to  Ref.~\cite{Wies89alc}
the width of the resonance is related to the spacing of the different levels at
the avoided crossing.  Therefore, we conclude that the resonance is very narrow,
and find good qualitative agreement with the parameters extracted from the
phase shift.  For the weaker potential ($V_0=2.0$, right panel of
Fig.~\ref{fig:En-2b-Dietz-A1p}), on the other hand, the avoided level crossings
are less sharp, pointing to a larger resonance width.  Along with the observed
sequence of plateaus at approximately $E=1.6$, we again find good qualitative
agreement with the phase-shift calculation.

For a more definite analysis, we determine the inflection 
points of the level curves with a plateau shape and interpret these as an 
estimate for the resonance energy.  For this extraction, we fit the 
coefficients $\{c_i\}$ of a polynomial,
\begin{equation}
 E(L) = \sum_{i=0}^{i_\text{max}} c_i L^i \,,
\label{eq:fitfunction}
\end{equation}
to the plateau region of each curve in Fig.~\ref{fig:En-2b-Dietz-A1p} and take
the position of the plateau inflection point as the resonance energy.  
In the plots, we indicate these points with crosses.  We vary the number of data
points taken into account for the fit by adjusting the lower
and upper boundary of the fit interval.  Furthermore, we vary $i_\text{max}$ in
Eq.~\eqref{eq:fitfunction} until we find the extracted resonance energy to be
independent of the order of the polynomial.  For $V_0=6.0$ and $V_0=2.0$ we obtain,
respectively, $E_R=2.98(3)$ and $E_R=1.63(3)$, where the quoted errors
correspond to the spread of the extracted inflection points from different
plateau curves.  This means that with the inflection-point method we obtain very
good agreement with the resonance positions from the phase-shift determination,
which justifies the use of this method for the resonance-energy
extraction.

At higher energies the spectra for both $V_0=6.0$ and $V_0=2.0$ exhibit less
pronounced avoided level crossings.  These structures, however, do not show
clear plateaus, instead varying strongly as a function of the box size.  Most
likely these finite-volume features correspond to the resonancelike jumps of
the phase shift at $E\sim 6-10$ for $V_0=6.0$ and $E\sim 3-7$ for $V_0=2.0$,
respectively, which may correspond to broader resonances.

Altogether, we have demonstrated here that the positions of narrow two-body
resonances can be extracted from finite-volume calculations with very good
quantitative agreement compared to other methods.

\section{Applications to three and four particles}
\label{sec:Res-34}

We now proceed to explore the method in the three- and four-body sector,
starting with bosonic (spin-0) particles.  Because these lack a spin degree of
freedom, we can quite easily achieve large DVR basis dimensions for these
systems, whereas fermionic systems are more computationally demanding.

\subsection{Three-body benchmark}

To verify our hypothesis that, analogously to the two-body case,
three-particle resonances appear as avoided level crossings in finite-volume
spectra, we start with three identical spin-0 bosons with mass $m=939.0~\MeV$
(mimicking nucleons) interacting \via the two-body potential,
\begin{equation}
 V(r) = V_0 \exp\biggl(-\Bigl(\frac{r}{R_0}\Bigr)^2\biggr)
 + V_1 \exp\biggl(-\Bigl(\frac{r-a}{R_1}\Bigr)^2\biggr) \,,
\label{eq:potblandon}
\end{equation}
where $V_0=-55~\MeV$, $V_1=1.5~\MeV$, $R_0=\sqrt{5}~\fm$, $R_1=10~\fm$, and
$a=5~\fm$.  This setup was studied in Ref.~\cite{Fedo03compscale}, where Faddeev
equations with complex scaling were used to calculate resonances, as well as in
Ref.~\cite{Blan07CTBR}, which used slow-variable discretization to extract
three-body resonance parameters.  The potential given in 
Eq.~\eqref{eq:potblandon} supports a two-body bound state (dimer) at 
$E=-6.76~\MeV$~\cite{Fedo03compscale} and a
three-boson bound state at $E=-37.35~\MeV$~\cite{Blan07CTBR}
(Ref.~\cite{Fedo03compscale} obtained $E=-37.22~\MeV$ for this state).  In
addition, it was found that there is a three-boson resonance at
$E_R=-5.31~\MeV$ with a half width of $0.12~\MeV$~\cite{Blan07CTBR}
($E_R=-5.96~\MeV$ and $\Gamma/2=0.40~\MeV$ according to 
Ref.~\cite{Fedo03compscale}), which decays into a dimer-particle state that is
overall lower in energy.

Using Eq.~\eqref{eq:potblandon} with our DVR method, we find $E=-6.756(1)$ and
$E=-37.30(5)$ for two and three bosons, respectively, in good agreement with the
results of Refs.~\cite{Blan07CTBR,Fedo03compscale}.  Note that bound-state
energies converge exponentially to the physical infinite-volume values as we
increase the box size $L$ (see, e.g., Refs.~\cite{Koen17VDNB,Meis153bb}).  In order to
look for the three-boson resonance, we study the positive-parity three-body 
spectrum as a function of $L$.  For  small box sizes around $L\sim 20~\fm$, we
find that $N=26$ DVR points is sufficient to obtain converged results.  For
large box size ($L\sim 40~\fm$), on the other hand, we performed calculations
using $N=30$.  The terms ``small'' and ``large'' here refer to the scale set by
the range of the interaction, which is quite sizable for the parameters given
below Eq.~\eqref{eq:potblandon}.

\begin{figure}[tbp]
\centering
\includegraphics[width=0.99\columnwidth]{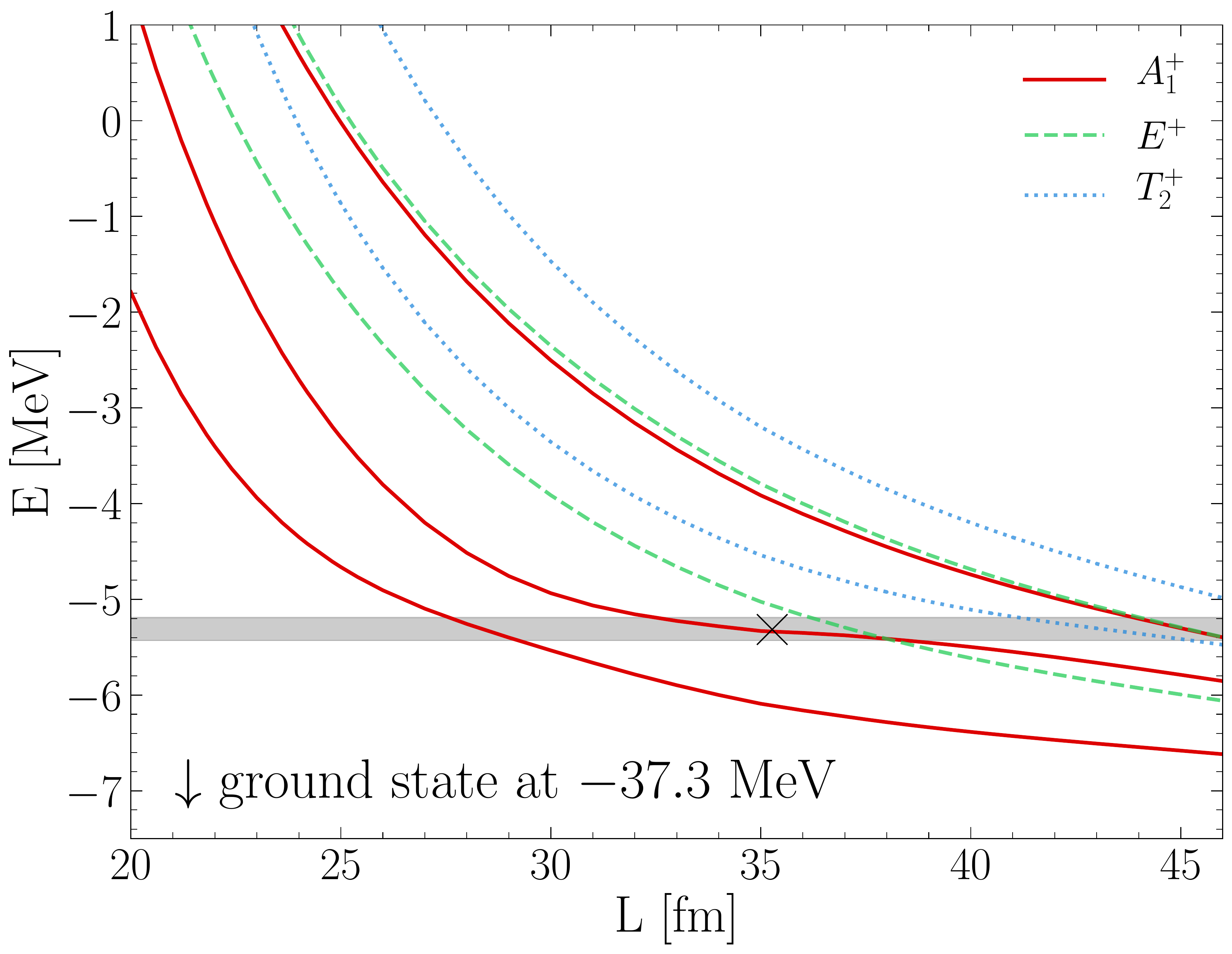}
\caption{Energy spectrum of three bosons in finite volume for different box
sizes $L$ interacting via the potential given in Eq.~\eqref{eq:potblandon}. 
States corresponding to the irreducible representation $A_1$ of the cubic 
symmetry group are shown as solid lines, whereas $E^+$ and $T_2^+$ states are
indicated as dashed and dotted lines, respectively.  The shaded area 
indicates the resonance position and width as calculated in 
Ref.~\cite{Blan07CTBR}, whereas the cross marks the inflection point 
used here to extract the resonance energy (see text).}
\label{fig:En-3b-Blandon-1-Pp}
\end{figure}

Our combined results are shown in Fig.~\ref{fig:En-3b-Blandon-1-Pp}, where we
also indicate the irreducible representations of the energy levels shown.  These
assignments were determined by running a set of cubic-projected calculations at
small volumes.  The levels corresponding to $A_1^+$ clearly show an avoided
crossing at about the expected resonance energy from Ref.~\cite{Blan07CTBR},
which is indicated in Fig.~\ref{fig:En-3b-Blandon-1-Pp} as a shaded horizontal
band, the width of which corresponds to $E_R\pm\Gamma/2$.  For the other
states (with quantum numbers $E^+$ and $T_2^+$) shown in the figure we do
not observe avoided crossings or plateaus.  At $L\sim 38~\fm$ there is an actual
crossing between $A_1^+$ and an $E^+$ levels.  This is not a very sharp avoided 
crossing because the participating levels belong to different 
cubic representations.

%%%%%%%%%%%%%%%%%%%%%%%%%%%%%%%%%%%%%%%%%%%%%%%%%%%%%%%%%%%%%%%%%%%%%%%%%%%%%%
%%%%%%%%%%%%%%%%%%%%%%%%%%%%%%%%%%%%%%%%%%%%%%%%%%%%%%%%%%%%%%%%%%%%%%%%%%%%%%

To extract the resonance energy from the spectrum shown  in
Fig.~\ref{fig:En-3b-Blandon-1-Pp} we proceed as described in
Sec.~\ref{sec:ResSig} and extract the  inflection points of the curves
corresponding to the $A_1^+$ states by fitting polynomials. For the first
excited state we find the fit to be quite sensitive to the number of data points
included in the fit, which reflects the fact that this level does not exhibit
a pronounced plateau.  For the second excited state, however, there is a
clearly visible plateau.  Applying our fit method to this state, we extract a
resonance energy $E_R = -5.32(1)~\MeV$.  This means that within the quoted
uncertainty, determined by varying the number of data points included in the fit
as well as the order of the fit polynomial, we obtain good agreement with the
resonance energy obtained in Ref.~\cite{Blan07CTBR}.  While a determination of
the resonance width is left for future work, we conclude from this result that
indeed finite-volume spectra can be used to reliably determine the existence and
energy of few-body resonances.

%%%%%%%%%%%%%%%%%%%%%%%%%%%%%%%%%%%%%%%%%%%%%%%%%%%%%%%%%%%%%%%%%%%%%%%%%%%%%%
\begin{figure}[btp]
\includegraphics[width=0.99\columnwidth]{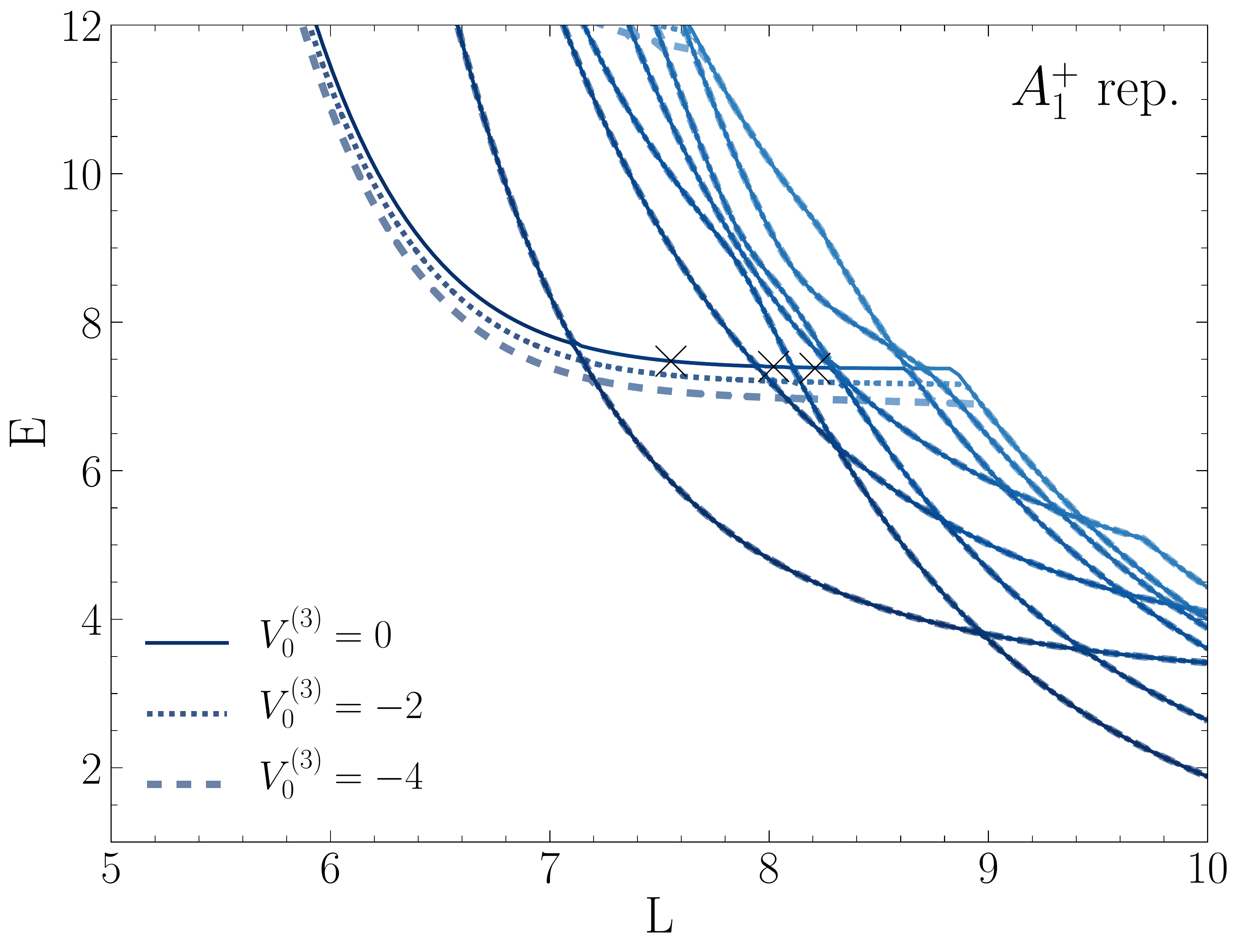}
\caption{Energy spectrum of three bosons in finite volume for different box
sizes $L$.  The solid lines shows the spectrum for three bosons interacting
purely via the shifted Gaussian potential given in Eq.~\eqref{eq:shiftedGauss}
with $V_0=6.0$ while the dashed and dotted lines show results with an additional
attractive three-body force as in Eq.~\eqref{eq:V3}. With increasing three-body
force, the avoided level crossing is shifted to lower energy, while the rest of
the spectrum remains unaffected.  For each choice of the three-body 
force, all crossings are avoided because the spectrum is fully projected on 
states with the same quantum numbers.  The crosses mark 
the inflection points used to extract the resonance energy (see text).}
\label{fig:En-d3n3b-N10-A1p}
\end{figure}
%%%%%%%%%%%%%%%%%%%%%%%%%%%%%%%%%%%%%%%%%%%%%%%%%%%%%%%%%%%%%%%%%%%%%%%%%%%%%%

\subsection{Shifted Gaussian potentials}
\label{sec:shiftedG}

\subsubsection{Three bosons}
\label{sec:shiftedG-3b}

%%%%%%%%%%%%%%%%%%%%%%%%%%%%%%%%%%%%%%%%%%%%%%%%%%%%%%%%%%%%%%%%%%%%%%%%%%%%%%
\begin{figure*}[bhtp]
\centering
\begin{minipage}{0.5\textwidth}
\centering
\includegraphics[width=0.99\columnwidth]{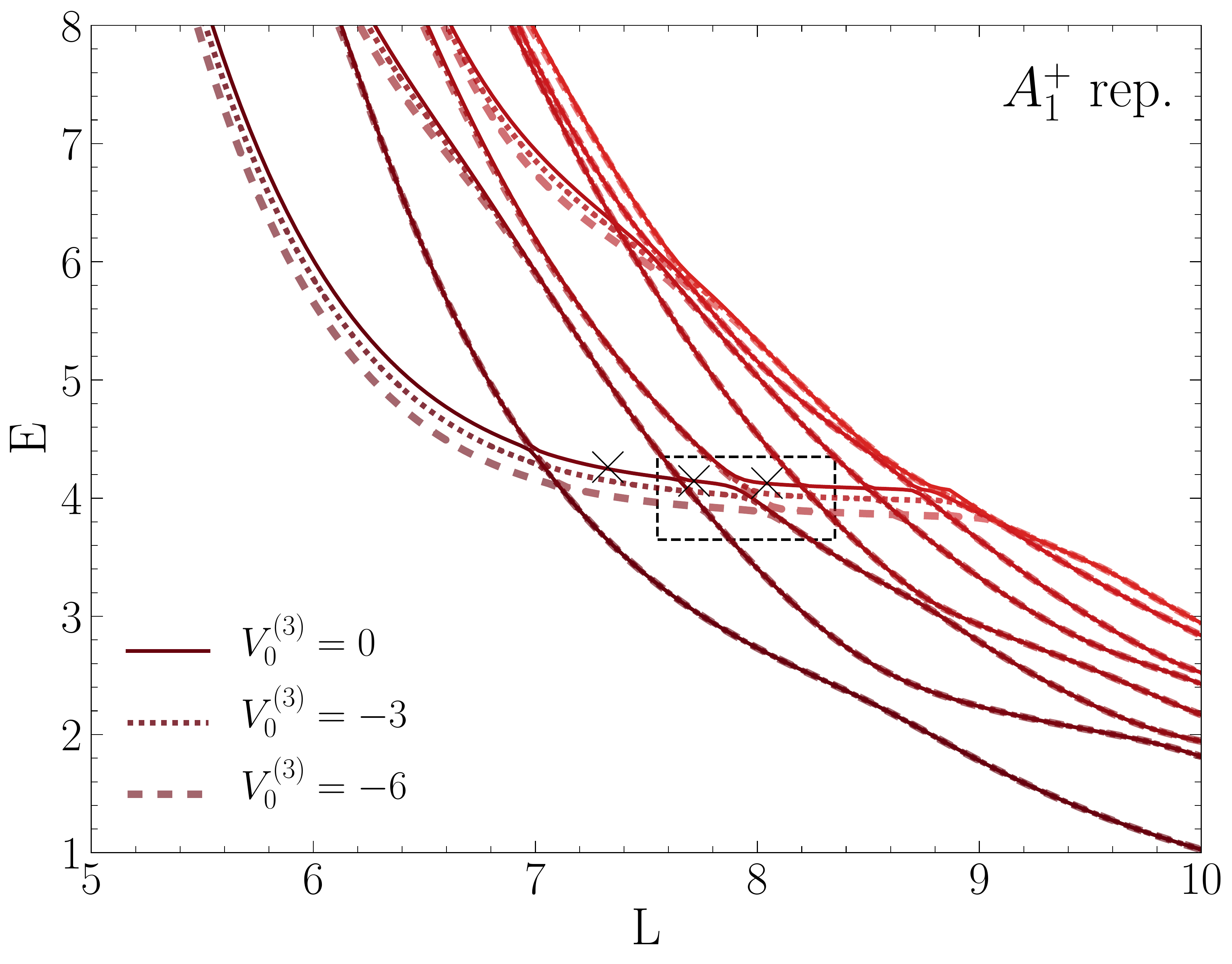}
\end{minipage}\begin{minipage}{0.5\textwidth}
\centering
\includegraphics[width=0.99\columnwidth]{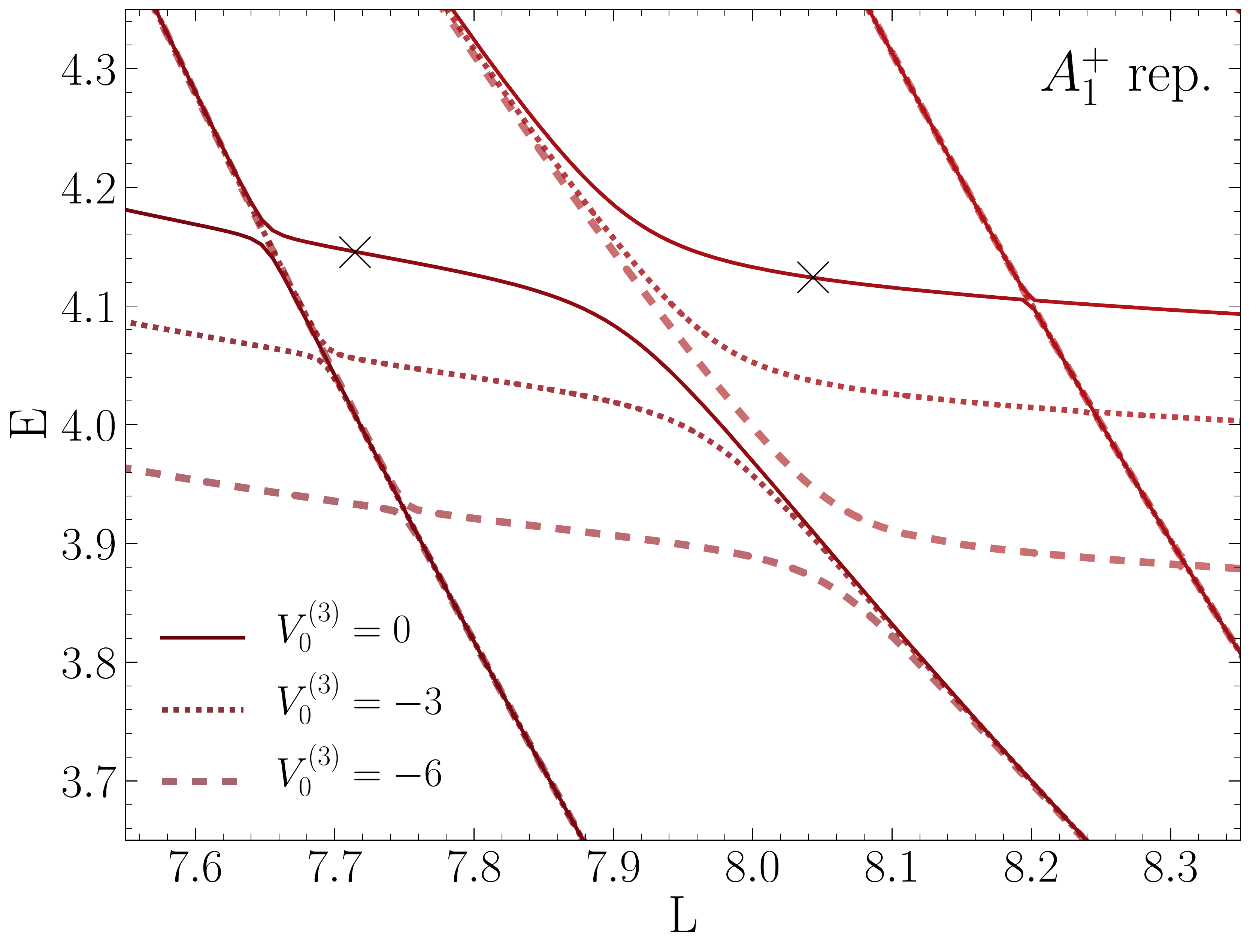}
\end{minipage}
\caption{Energy spectrum of three bosons in finite volume for different box
sizes $L$.  The solid line shows the spectrum for three bosons  interacting
purely via the shifted Gaussian potential given in Eq.~\eqref{eq:shiftedGauss}
with $V_0=2.0$ while the dashed and dotted lines show results with an additional
attractive three-body force as in Eq.~\eqref{eq:V3}. With increasing three-body
force the avoided level crossing is shifted to lower energy, while the rest of
the spectrum remains unaffected.  The dashed rectangle in the left panel
indicates the zoomed region shown in the right panel.  For each choice 
of the three-body force, all crossings are avoided because the spectrum is 
fully projected on states with the same quantum numbers.  The crosses 
mark the inflection points used to extract the resonance energy (see text).}
\label{fig:En-Dietz2p0-N10-A1p}
\end{figure*}
%%%%%%%%%%%%%%%%%%%%%%%%%%%%%%%%%%%%%%%%%%%%%%%%%%%%%%%%%%%%%%%%%%%%%%%%%%%%%%

Having established the validity of the finite-volume method to extract
three-body resonances, we now go back to the shifted Gaussian
potential given in Eq.~\eqref{eq:shiftedGauss} which was used in
Sec.~\ref{sec:ResSig} to study two-body resonances.  Starting again with the
stronger barrier, ($V_0=6.0$), we consider the $A_1^+$ spectrum for three
bosons, calculated with $N=10$ DVR points and shown in
Fig.~\ref{fig:En-d3n3b-N10-A1p} as solid lines.  We observe a large number of
avoided crossings at $E\sim 7.4$ as the box size $L$ is varied, producing
together an almost horizontal plateau region.  Using the same inflection-point
method as discussed above, we extract $E = 7.42(6)$ as a potential resonance
energy.  In addition to this, there are several avoided crossings at lower
energies that have a significant slope with respect to changes in the box size,
which we interpret as two-body resonances (known from Sec.~\ref{sec:ResSig} to
exist at $E_{R} \sim 3.0$ for this potential) embedded into the three-body
spectrum.  To test this hypothesis we repeat the calculation with
an added short-range three-body force,
\begin{multline}
 V_{3}(x_1,x_2,x_{12}) = V_0^{(3)}
 \exp\Biggl({-}\biggl(\frac{x_1}{R_0^{(3)}}\biggr)^2\Biggr) \\
 \null\times\exp\Biggl(-\biggl(\frac{x_2}{R_0^{(3)}}\biggr)^2\Biggr)
 \exp\Biggl({-}\biggl(\frac{x_{12}}{R_0^{(3)}}\biggr)^2\Biggr) \,,
\label{eq:V3}
\end{multline}
where $x_{12}=|\vecx_1-\vecx_2|$ and $R_0^{(3)}=1.0$ and varying strength
$V_0^{(3)}$.  Choosing a set of negative values for $V_0^{(3)}$ we find in
Fig.~\ref{fig:En-d3n3b-N10-A1p} that the lower avoided crossings (and in fact
most of the $L$-dependent spectrum) remain unaffected, whereas the upper plateau
set is moved downwards as $V_0^{(3)}$ is made more negative.

Since the range $R_0^{(3)}=1.0$ was chosen small (compared to the
box sizes considered), we expect it to primarily affect states that are
localized in the sense that their wave function is confined to a relatively
small region in the finite volume.  Interpreting a resonance as a nearly bound
state, its wave function should satisfy this criterion in the finite volume.  On
the other hand, scattering states or states where only two particles are bound
or resonant are expected to have a large spatial extent.  Based on this
intuitive picture, we interpret the action of the three-body force as
confirmation that indeed we have a genuine (because the potential we used does
not support any bound states) three-boson resonance state at $E = 7.42(6)$.

Similar to the two-body spectrum shown in the left panel of
Fig.~\ref{fig:En-2b-Dietz-A1p} we find that Eq.~\eqref{eq:shiftedGauss} with
$V_0=6.0$ generates very sharp features in the three-boson spectrum so that
even though we used a fine $L$ grid to generate Fig.~\ref{fig:En-d3n3b-N10-A1p}
it is difficult to exclude that some crossings might not actually be avoided
crossings.  However, we observe the exact same qualitative behavior for the
potential given in Eq.~\eqref{eq:shiftedGauss} with $V_0=2.0$, only that in 
this case the avoided crossings are broader and easily identified.  From the
spectrum, shown in Fig.~\ref{fig:En-Dietz2p0-N10-A1p}, we extract 
$E=4.18(8)$ as the three-boson resonance energy for this case.

\subsubsection{Four bosons}
\label{sec:shiftedG-4b}

\begin{figure*}[tb]
\centering
\begin{minipage}{0.5\textwidth}
\centering
\includegraphics[width=0.99\columnwidth]{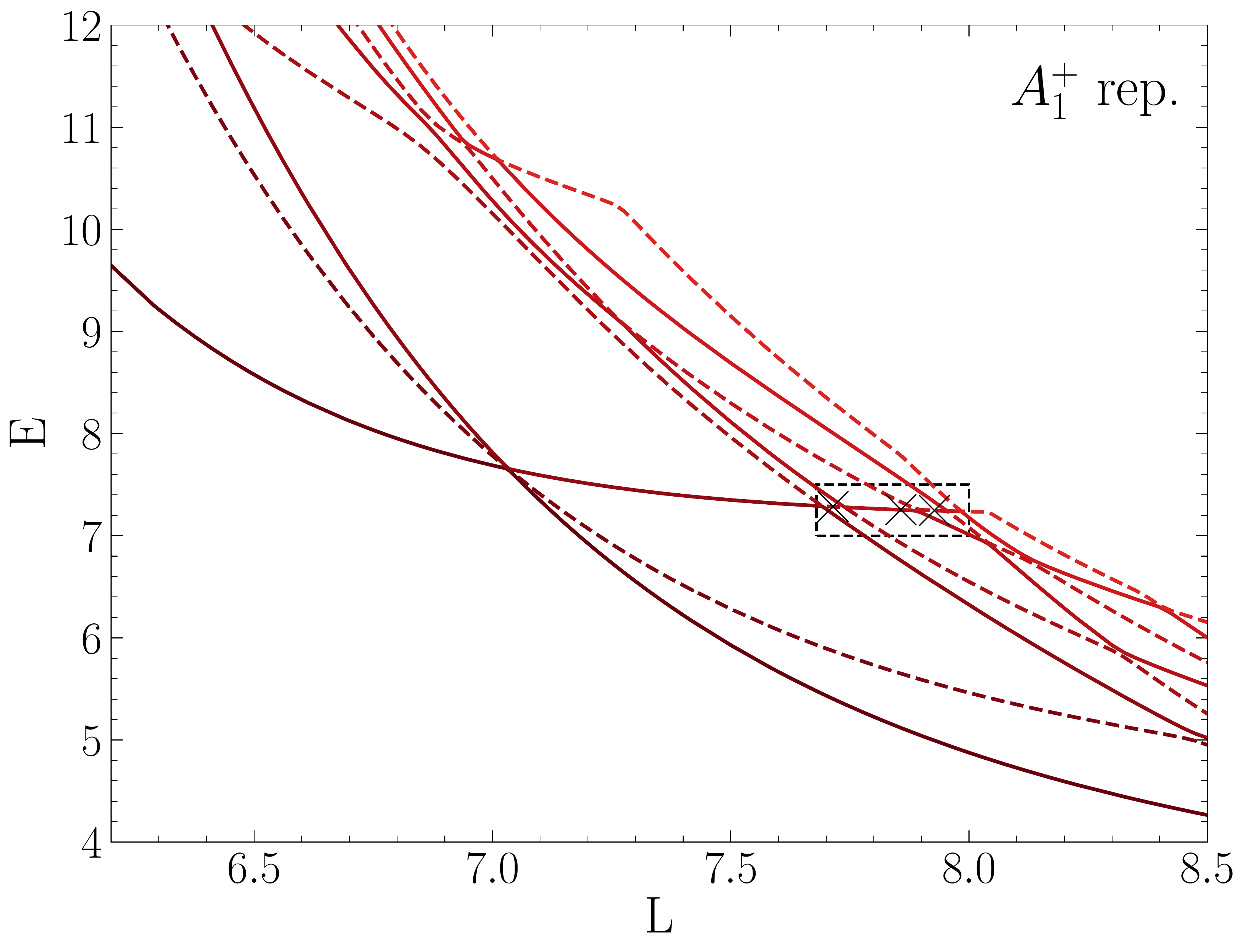}
\end{minipage}\begin{minipage}{0.5\textwidth}
\centering
\includegraphics[width=0.99\columnwidth]{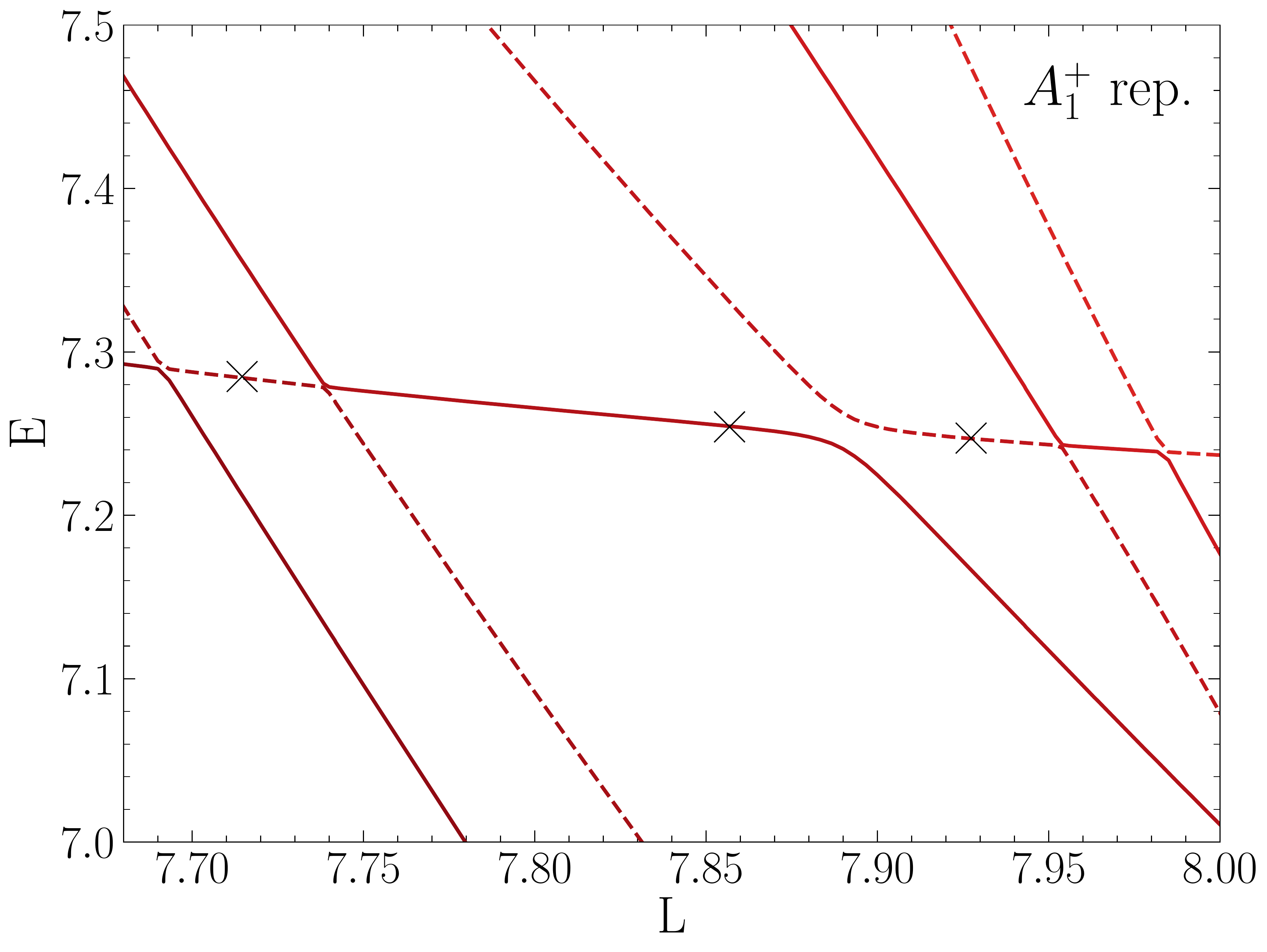}
\end{minipage}
\caption{Energy spectrum of four bosons in finite volume for
different box sizes $L$ interacting \via the shifted Gaussian potential given in
Eq.~\eqref{eq:shiftedGauss} with $V_0=2.0$.  The dashed rectangle in the left
panel indicates the zoomed region shown in the right panel.  All crossings are
avoided because the spectrum is fully projected on states with the 
same quantum numbers.  The crosses mark the inflection points used to extract 
the resonance energy (see text).}
\label{fig:En-4b-Dietz2p0-A1p-N8}
\end{figure*}

Looking next at four bosons, we find a very similar
picture.  As shown in Fig.~\ref{fig:En-4b-Dietz2p0-A1p-N8} for the shifted
Gaussian potential given in Eq.~\eqref{eq:shiftedGauss} with $V_0=2.0$, the
$L$-dependent $A_1^+$ four-boson spectrum (calculated with $N=8$ DVR points in
this case) shows a large number of avoided level crossings that give rise to
plateaus with different slopes.  Interpreting the nearly horizontal set of
avoided crossings as a possible four-boson resonance, we extract its energy as
$E=7.26(2)$ with the inflection-point method.  The more tilted sets of
avoided crossings at lower energies most likely correspond to two- and
three-boson resonance states embedded in the four-boson spectrum.

%%%%%%%%%%%%%%%%%%%%%%%%%%%%%%%%%%%%%%%%%%%%%%%%%%%%%%%%%%%%%%%%%%%%%%%%%%%%%%
%%%%%%%%%%%%%%%%%%%%%%%%%%%%%%%%%%%%%%%%%%%%%%%%%%%%%%%%%%%%%%%%%%%%%%%%%%%%%%

\subsubsection{Three fermions}
\label{sec:shiftedG-3f}

To conclude our survey, we now turn to fermionic systems.  As the
additional spin degree of freedom (we consider here identical spin-$1/2$
particles) increases the DVR basis size [see discussion below Eq.~\eqref{eq:s}],
these calculations are more computationally demanding, but we can still achieve
well-converged results for the shifted Gaussian potential given in
Eq.~\eqref{eq:shiftedGauss}.  Before we turn to the three-body sector, we
note that the results of Sec.~\ref{sec:ResSig} remain correct when we assume the
two fermions to be in the channel with total spin $S=0$.  In this case, the
spin part of the wave function is antisymmetric and the spatial part has to be
even under exchange.  Because the latter corresponds to the bosonic case with
positive parity, we conclude that for two spin-$1/2$ fermions the two-body
potential given in Eq.~\eqref{eq:shiftedGauss} has a resonance state at
$E_{R}\sim1.6$ for $V_0=2.0$.

%%%%%%%%%%%%%%%%%%%%%%%%%%%%%%%%%%%%%%%%%%%%%%%%%%%%%%%%%%%%%%%%%%%%%%%%%%%%%%
\begin{figure}[bt]
\centering
\includegraphics[width=0.99\columnwidth]{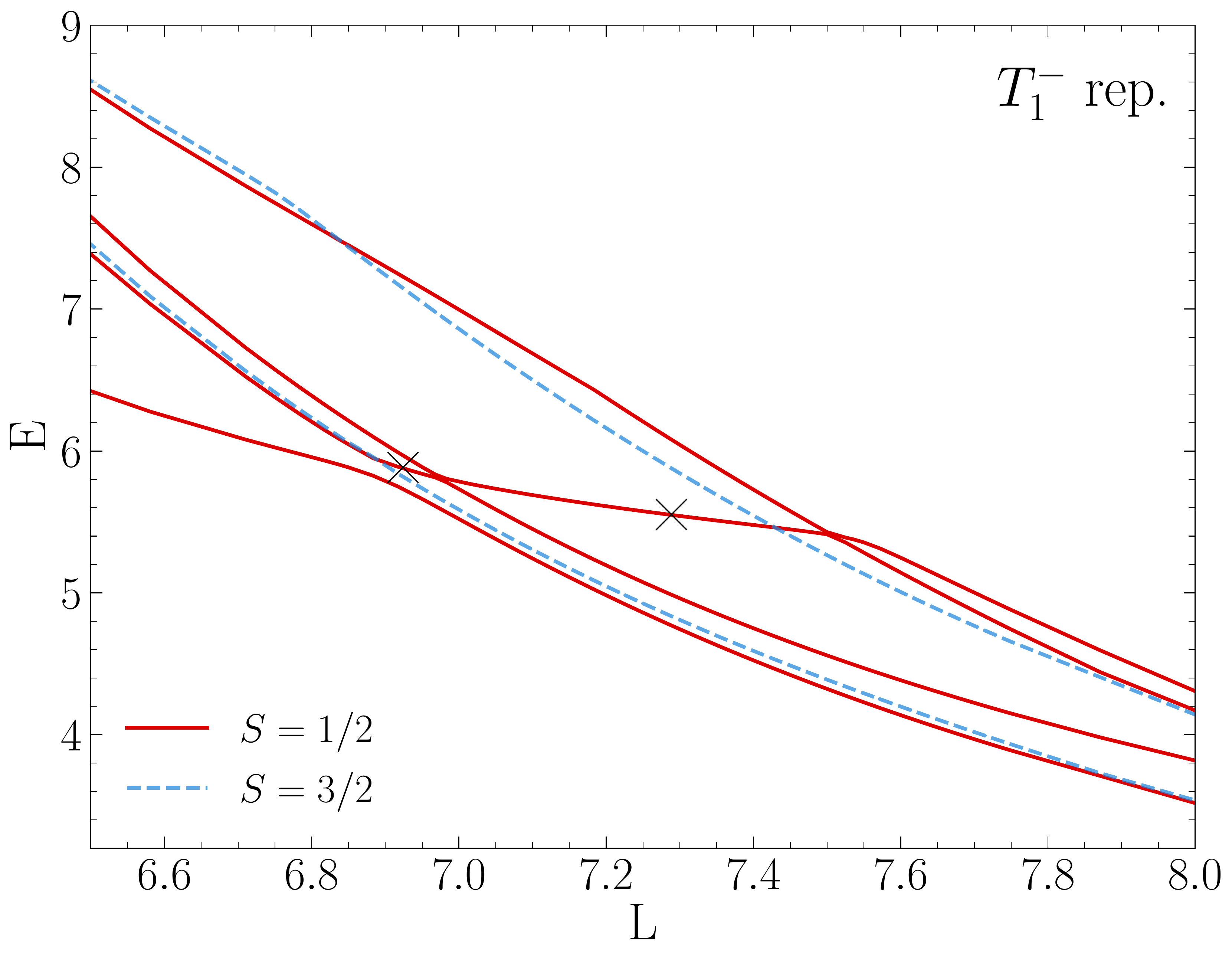}
\caption{Negative-parity energy spectrum of three fermions in finite volume for
different box sizes $L$ interacting \via the shifted Gaussian potential given in
Eq.~\eqref{eq:shiftedGauss} with $V_0=2.0$.  All levels shown in the plot were
found to belong to the $T_1^-$ cubic representation by performing fully
projected calculations at selected volumes.  Results are shown in the spin
$S=1/2$ and $S=3/2$ channels.  The crosses mark the inflection points used to
extract the resonance energy (see text).}
\label{fig:En-3f-Dietz2p0-Pm}
\end{figure}
%%%%%%%%%%%%%%%%%%%%%%%%%%%%%%%%%%%%%%%%%%%%%%%%%%%%%%%%%%%%%%%%%%%%%%%%%%%%%%

For three fermions, on the other hand, the situation is more involved because
the overall antisymmetry of the wave function can be realized \via different
combinations of spin and spatial parts.  Indeed, we find the finite-volume
spectrum to look different from the bosonic case.  For negative parity, we find
the six lowest levels, shown in Fig.~\ref{fig:En-3f-Dietz2p0-Pm}, to all belong
to the $T_1^-$ cubic representation, which in this case we determined by running
calculations with full cubic projections at selected volumes while otherwise
only restricting the overall parity.  Because the interaction we consider here
is spin independent, total angular momentum $l$ and spin $S$ are separately
good quantum numbers in infinite volume, and in the finite volume we
likewise have $\Gamma$ and $S$ as good quantum numbers.  The latter,
which can be $S=1/2$ or $S=3/2$ for three spin-$1/2$ fermions, we determine by
running calculations with fixed spin $z$-component at selected volumes, which
can be realized by restricting the set of DVR basis states.  Because $S=3/2$
states show up with both $S_z=3/2$ and $S_z=1/2$, whereas $S=1/2$ states are
absent for $S_z=3/2$, we infer that four of the six levels shown in
Fig.~\ref{fig:En-3f-Dietz2p0-Pm} have $S=1/2$, whereas the other two (given by
the dashed lines in Fig.~\ref{fig:En-3f-Dietz2p0-Pm}) have $S=3/2$.

For $S=1/2$ we observe a sequence of three avoided level crossings.  Within
this sequence there is a drift towards lower energies as $L$ increases, the
magnitude of which is comparable to what we observe also for the three-boson
spectra analyzed in Sec.~\ref{sec:shiftedG-3b} for the state that we concluded
to correspond to an actual three-body resonance (based on varying the
three-body force).  We thus conclude that this effect is likely a residual
volume dependence of an actual resonance state also in this case.  With this
interpretation, we extract a resonance energy $E_R=5.7(2)$ from the spectrum
shown in Fig.~\ref{fig:En-3f-Dietz2p0-Pm} with our inflection-point method.

\section{Summary and outlook}
\label{sec:Conclusion}

We established the method of analyzing few-body energy spectra in finite
periodic boxes to extract three- and four-particle resonance energies.
Our approach relies on the observation of avoided level crossings and/or
plateaus in the spectra considered as a function of the box size.  Observing
such features in few-body spectra and showing that they can be used to find and
analyze resonance states, thus generalizing the method introduced in
Ref.~\cite{Wies89alc} for two-body systems, is the central result of this work.

To calculate the finite-volume spectra, which were then used for the
resonance identification, we used a DVR basis based on plane-wave states in
relative coordinates.  Resonance features are expected for finite-volume
energies corresponding to scattering states in infinite volume.  Unlike bound
states, the energies of which converge exponentially with the box size $L$,
finite-volume scattering states have a power-law dependence on $L$
(away from regions with avoided crossing).  Looking at low-energy resonances
therefore requires going to volumes that are sufficiently large for the relevant
levels to come down to the energy range of interest.  Because calculations in
this regime typically require large DVR basis sizes and become computationally
very demanding, we have developed a numerical framework to run the calculations 
on high-performance computing clusters when necessary.  We have furthermore 
extended the formalism to include the symmetrization (antisymmetrization) to 
study bosonic (fermionic) systems, as well as for projecting onto the subspaces 
belonging to parity eigenstates and to the different irreducible representations 
of the cubic symmetry group.  The latter allows us to determine the 
finite-volume quantum numbers of the resonance states that we find.

After testing our method in the two-body sector, where we verified the
existence of resonances by looking at characteristic jumps in the scattering
phase shifts as well as by looking for $S$-matrix poles on the second energy
sheet, we studied three- and four-body systems with different potentials.
First, we used a model potential known to generate a three-boson resonance that
decays into a lower lying two-body bound state and a free particle.  For this
system, the resonance parameters were extracted previously based on different
methods~\cite{Blan07CTBR,Fedo03compscale}.  Our results clearly show an
avoided level crossing in the corresponding finite-volume spectrum and we find
good agreement with the resonance energy of Ref.~\cite{Blan07CTBR}, which we extracted from the inflection
points of the volume-dependent energy levels.

Taking this agreement as confirmation that our method works both qualitatively
and quantitatively, we used shifted Gaussian potentials (with the same
parameters known to generate two-body resonances) in the three- and four-body
sector.  Studying the three-boson finite-volume spectrum, we showed that an
additional short-range three-body force can be used to move avoided crossings
forming a plateau region whereas other avoided crossings remain unchanged.  We
interpret this as confirmation that the observed plateau region indeed
corresponds to a three-body resonance (with a spatially localized wave function
so that it ``feels'' the three-body forces), whereas the other levels likely
correspond to two-body resonances plus a third particle.  For the same shifted
Gaussian potential we were also able to observe avoided crossings for three
fermions and four bosons, from which we extracted resonance energies \via the
inflection-point method.  Based on these findings, we conclude that our method
can be used to search for possible three- and four-neutron resonances in
future work.

\begin{acknowledgments}
We thank Akaki Rusetsky for valuable discussions.  This work was supported by
the ERC Grant No.\ 307986 STRONGINT, the Deutsche Forschungsgesellschaft (DFG)
under Grant SFB 1245, and the BMBF under Contract No.~05P15RDFN1.  The numerical
computations were performed on the Lichtenberg high performance computer of the 
TU Darmstadt and at the Jülich Supercomputing Center.
\end{acknowledgments}

\appendix

\section{Cubic symmetry group}
\label{app:cubic}

In this section, we briefly discuss how the projector on the irreducible 
representations of the cubic symmetry group in Eq.~\eqref{eq:P_Gamma} is 
constructed.  For each element of the cubic group $R\in\OO$, the
realization $D_{n}(R)$ used in Eq.~\eqref{eq:P_Gamma} is given by a permutation 
and/or inversion of the components $c=1,2,3$ of each relative coordinate 
$\vecx_i$ (simultaneously for all $i=1,\ldots,n-1$).
%
%%%%%%%%%%%%%%%%%%%%%%%%%%%%%%%%%%%%%%%%%%%%%%%%%%%%%%%%%%%%%%%%%%%%%%%%%%%%%%%%
\begin{table}[t!]
\begin{tabular}{c|c|rrr||c|c|rrr}
 Index & Class & \multicolumn{3}{c||}{$D_n(R)$} &
 Index & Class & \multicolumn{3}{c}{$D_n(R)$} \\
 \hline
  1 & $I$ & $1$ & $2$ & $3$ &          13 & $6C_4$ & $2$ & ${-}1$ & 3 \\
  2 & $3C_2$ & ${-}1$ & ${-}2$ & $3$ & 14 & & ${-}2$ & 1 & 3 \\
  3 & & ${-}1$ & $2$ & ${-}3$ &        15 & & $3$ & $2$ & ${-}1$ \\
  4 & & $1$ & ${-}2$ & ${-}3$ &        16 & & ${-}3$ & $2$ & $1$ \\
  5 & $8C_3$ & $3$ & $1$ & $2$ &       17 & & $1$ & ${-}3$ & $2$ \\
  6 & & $2$ & $3$ & $1$ &              18 & & $1$ & $3$ & ${-}2$ \\
  7 & & ${-}2$ & $3$ & ${-}1$ &        19 & $6C_2'$ & $2$ & $1$ & ${-}3$ \\
  8 & & ${-}3$ & ${-}1$ & $2$ &        20 & & ${-}2$ & ${-}1$ & ${-}3$ \\
  9 & & $2$ & ${-}3$ & ${-}1$ &        21 & & $3$ & ${-}2$ & $1$ \\
 10 & & ${-}3$ & $1$ & ${-}2$ &        22 & & ${-}3$ & ${-}2$ & ${-}1$ \\
 11 & & ${-}2$ & ${-}3$ & $1$ &        23 & & ${-}1$ & ${-}3$ & ${-}2$ \\
 12 & & $3$ & ${-}1$ & ${-}2$ &        24 & & ${-}1$ & $3$ & $2$
\end{tabular}
\caption{Realization of the 24 cubic rotations acting on a coordinate tuple in
symbolic notation (see text).  The second column indicates the conjugacy class
of the rotation.}
\label{tab:O}
\end{table}
%%%%%%%%%%%%%%%%%%%%%%%%%%%%%%%%%%%%%%%%%%%%%%%%%%%%%%%%%%%%%%%%%%%%%%%%%%%%%%%%
\begin{table}[b!]
\begin{center}
\begin{tabular}{c|ccccc|ccccc}
$l$&$A^+_1$&$A^+_2$&$E^+$&$T^+_1$&$T^+_2$&$A^-_1$&$A^-_2$&$E^-$&$T^-_1$&$T^-_2$
\\
\hline\hline
$0$&1&&&&\\
$1$&&&&&&&&&1\\
$2$&&&1&&1&\\
$3$&&&&&&&1&&1&1\\
$4$&1&&1&1&1&\\
$5$&&&&&&&&1&2&1\\
$6$&1&1&1&1&2&\\
$7$&&&&&&&1&1&2&2\\
$8$&1&&2&2&2&\\
$9$&&&&&&1&1&1&3&2\\
$10$&1&1&2&2&3&
\end{tabular}
\caption{Decomposition of the irreducible representations of the rotational
symmetry group $SO(3)$ into irreducible representations of the cubic symmetry
group $\mathcal{O}$; reproduced in part from Ref.~\cite{Dress02Group}.}
\label{tab:cubicL}
\end{center}
\end{table}
%%%%%%%%%%%%%%%%%%%%%%%%%%%%%%%%%%%%%%%%%%%%%%%%%%%%%%%%%%%%%%%%%%%%%%%%%%%%%%%%
%
In Table~\ref{tab:O} we 
show these operations, where the notation gives the result of operating on a 
tuple $(k_{i,1},k_{i,2},k_{i,3})$ in a short-hand form, \eg, the rotation with 
index 7 transforms a tuple to $(-k_{i,2},k_{i,3},-k_{i,1})$.  It is understood 
that, as discussed in Sec.~\ref{sec:AntiSymm-Parity}, each transformed index 
$k_{i,c}$ is wrapped back into the interval ${-}N/2,\ldots,N/2-1$, if necessary.

Cubic symmetry commutes with parity as well as permutation symmetry, so for
both bosonic and fermionic systems we end up with multiplets of the irreducible
representations $\Gamma = A_1^\pm$, $A_2^\pm$, $E^\pm$, $T_1^\pm$, and $T_2^\pm$,
where the superscript indicates the parity. As already mentioned above, the
irreducible representation of the full rotational group $SO(3)$ is reducible
with respect to the cubic group. A basis for the irreducible representation of
$SO(3)$ is given by the angular momentum multiplets, \ie, spherical harmonics
$Y_{lm}$, labeled by the angular momentum quantum number $l$ and its projection
$m$. The numerical values in Table~\ref{tab:cubicL} yield the multiplicity of
the cubic irreducible representations in the decomposition of a given angular
momentum multiplet.  $l=0$ and $l=1$ contribute only to $A^+_1$ and $T^-_1$,
respectively, meaning that an $S$-wave state is mapped solely onto the single
$A_1^+$ state, while a $P$-wave state maps onto the three $T^-_1$ states in
finite volume.  A $D$-wave state with its five projections $m=0,\pm 
1,\pm 2$ is decomposed into the two $E^+$ and three $T_2^+$ states.

To conclude this section, we note that in the case of spin-dependent
interactions, total angular momentum $J$ instead of $l$ is the relevant good
quantum number in the infinite volume.  For example, in the case of spin-$1/2$
fermions, one has to consider $SU(2)$ broken down to the double cover ${}^2\OO$
of the cubic group, giving three additional irreducible representations that
receive contributions from half-integer $J$ states.  For details, see
Ref.~\cite{John82AML}.

\end{document}